\documentclass[sn-mathphys,Numbered]{sn-jnl}% Math and Physical Sciences Reference Style
\usepackage{graphicx}%
\usepackage{multirow}%
\usepackage{amsmath,amssymb,amsfonts}%
\usepackage{amsthm}%
\usepackage{mathrsfs}%
\usepackage[title]{appendix}%
\usepackage{xcolor}%
\usepackage{textcomp}%
\usepackage{manyfoot}%
\usepackage{booktabs}%
\usepackage{algorithm}%
\usepackage{algorithmicx}%
\usepackage{algpseudocode}%
\usepackage{listings}%
\usepackage{txfonts}%
\usepackage{lineno}%
\usepackage{comment}%
\usepackage{slashed}%
\usepackage[T1]{fontenc}%
\begin{document}

\title[Article Title]{On the synergic approach toward the experimental realization of interesting fundamental science within the framework of relativistic flying mirror concept}


\author*[1]{\fnm{Tae Moon} \sur{Jeong}}\email{taemoon.jeong@eli-beams.eu}

\author[1,2]{\fnm{Sergei V.} \sur{Bulanov}}\email{sergei.bulanov@eli-beams.eu}

\author[1]{\fnm{Petr} \sur{Valenta}}\email{petr.valenta@eli-beams.eu}

\author[1]{\fnm{Prokopis} \sur{Hadjisolomou}}\email{prokopis.hadjisolomou@eli-beams.eu}

\affil*[1]{\orgdiv{ELI-Beamlines}, \orgname{ELI-ERIC}, \orgaddress{\street{Za Radnicí 835}, \city{Dolní Břežany}, \postcode{25241}, \country{Czech Republic}}}

\affil[2]{\orgdiv{Kansai Photon Research Institute}, \orgname{National Institutes for Quantum and Radiological Science and Technology}, \orgaddress{\street{8-1-7 Umemidai}, \city{Kizugawa-shi}, \postcode{619-0215}, \state{Kyoto}, \country{Japan}}}


\abstract{The relativistic flying parabolic mirror can provide a higher laser intensity than the intensity a current laser system can reach via the optical-focusing scheme. A weakly-relativistic laser intensity (1.8 \unboldmath$\times$ 10$^{17}$ W/cm$^2$, $\eta = 0.29$) can be intensified up to a super-strong intensity of >1$\times$10$^{27}$ W/cm$^2$ ($\eta \approx 2.2 \times 10^4$) by the relativistic flying mirror. Such a super-strong field can be applied to study the strong-field quantum electrodynamics in perturbative and non-perturbative regimes. In this review, the analytic derivations on the field strength and distribution obtained by the ideal relativistic flying parabolic mirror have been shown under the 4$\pi$-spherical focusing approach. The quantum nonlinearity parameter is calculated when such a super-strong field collides with the high-energy $\gamma$-photons. The peak quantum nonlinearity parameter reaches above 1600 when the 1-GeV $\gamma$-photon collides with a super-strong laser field reflected and focused by the relativistic flying mirror driven by a 10 PW laser pulse.}

\keywords{Relativistic flying mirror, High-power laser, Critical field strength, Photon-gamma photon collision, Quantum nonlinearity parameter}



\maketitle

\section{Introduction}\label{sec1}

For the last three decades, the chirped pulse amplification (CPA) technique \cite{Strickland} has made tremendous progress in the femtosecond (fs) high-power laser development. The peak power of the laser pulse exceeds 10 PW \cite{Tanaka} and a laser intensity of > 10$^{23}$ W/cm$^2$ has been experimentally demonstrated by focusing a 4 PW laser pulse with an f/1.3 off-axis parabolic (OAP) mirror \cite{Yoon}. An excellent overview for the current and upcoming high-power laser systems around the world can be found in \cite{Danson}. The strong field strength (>GV/mm, 10$^9$ V/mm) formed by such a high-power laser pulse makes electrons relativistic in a short distance ($\sim$mm), helping ones realize the idea of a compact laser-driven charged particle (electron/proton/ions) accelerator \cite{Esarey_LEA,Daido,Macchi}. A dimensionless parameter, $\eta$, defined as,
%%%% Equation 1-1
\begin{equation} \label{eq1-1}
\eta = \frac{eE}{mc \omega}.     
\end{equation} 
is a measure of the field strength and plays a key role in these cases. Here, $e$ is the elementary charge, $E$ the electric field strength, $m$ the electron rest mass, $c$ the speed of light, and $\omega$ the angular frequency of the electromagnetic (EM) wave. The dimensionless parameter means the ratio between the energy ($e E \lambda$) gained by an electron from the EM wave in one wavelength ($\lambda$) and the rest-mass energy ($mc^2$) of the electron. The fundamental physics with such a strong field strength ($\eta > 1$) has been discussed under the context of relativistic optics in \cite{Mourou}. 

Despite the strong field strength can exceed a field strength of TV/mm (10$^{12}$ V/mm) with the currently available laser power, it is still three orders lower than the critical field strength, $E_c (= m^2 c^3 / e \hbar) \sim 1.3$ PV/mm, required for the spontaneous electron-positron (ep) pair creation (Schwinger pair creation) \cite{Schwinger}. However, a focused laser field can provide a strong background field to relativistic electrons, and the relativistic electron experiences an effective field strength enhanced by the Lorentz $\gamma$-factor. In this case, the photon emission by the relativistic electron is described by another dimensionless parameter,
%%%% Equation 1-2
\begin{equation} \label{eq1-2}
\chi_e = \frac{k_l \cdot q}{m^2} \eta, \quad \textrm{or} \quad \chi_e = \gamma \frac{ E}{E_c} \left( 1 - \cos \theta \right),
\end{equation}
which is known as the quantum nonlinearity parameter, $\chi_e$ \cite{Ritus,Harvey,Heinzl,Piazza,Mackenroth,Ridgers,Dinu,Tang}. In Eq. \eqref{eq1-2}, $k_l$ stands for the four-wavevector of the EM radiation (background laser field), $q$ the four-momentum of an electron, respectively. The polar angle, $\theta$, is defined against the laser propagation direction, so for the co-propagating case $\theta = 0$ and for the counter-propagating case $\theta = \pi$. The ep pair creation by the multi-photon Breit-Wheeler process \cite{Ritus,Reiss,Bell,Sokolov,Piazza_PP,Blackburn} can be described by using another quantum nonlinearity parameter defined as,
%%%% Equation 1-3
\begin{equation} \label{eq1-3}
\chi_{\gamma} = \frac{k_l \cdot k_{\gamma}}{m^2} \eta, \quad \textrm{or} \quad \chi_{\gamma} = \frac{\hbar \omega_{\gamma}}{m c^2} \frac{ E}{E_c} \left( 1 - \cos \theta \right),
\end{equation}
where $k_{\gamma}$ and $\omega_{\gamma}$ are the four-wavevector and the angular frequency of high energy photon, respectively. Through multi-photon absorption processes, the strong background field relaxes the physical conditions (i.e., laser intensity) required for quantum electrodynamic processes. These are now discussed under the context of strong-field quantum electrodynamics (SF-QED) \cite{Gonoskov_RMP}. It is known that $\alpha \chi^{2/3}$ is the expansion parameter for the perturbation theory in SF-QED when $\alpha \chi^{2/3} \ll 1$ \cite{Narozhny}, where $\alpha$ is the fine structure constant defined as $e^2 / 4 \pi \varepsilon_0 \hbar c$. 

The seminal experiment of the E144 experiment in SLAC utilizes $\chi_e$ and $\chi_{\gamma}$ for ep pair creation through multi-photon Compton scattering and multi-photon Breit-Wheeler processes \cite{Burke,Bamber}. Recent proposals on SF-QED experiments are based on this approach \cite{Yakimenko_A, Abramowicz}. In particular, the European collaboration \cite{Abramowicz} aims to perform precision measurements that can investigate the transition into the non-perturbative regime with elaborate diagnostics as well as a powerful detection system. On the other side, owing to the advance in the laser-driven electron acceleration \cite{Gonsalves}, the "all-optical" scheme is actively discussed for the SF-QED experiment \cite{SVBulanov_AO, TaPhuoc, Lobet, Cole, Poder, Blackburn1, MacLeod}. The intrinsic synchronization and the improved spatial overlap between the electron beam and the intense laser pulse might enable to explore the transition from the classical to the quantum radiation reaction regimes. 

Besides the above effort, the formation of an ultrastrong field is still of fundamental interest in the SF-QED. The relativistic flying parabolic mirror is proposed to further intensify a focused EM radiation beyond the optical focusing scheme \cite{SVBulanov_RFM,quéré}. The relativistic flying mirror is one of the plasma optics which generate bright high-energy photons via laser-matter interactions \cite{Bulanov1994, Lichters, Baeva, Dromey, Thaury, Pirozhkov_RFM, Kando_2007, Kando, Teubner, Gonoskov_RM, Jeong_RHHG, Kahaly, Debayle, VINCENTI}. Under specific laser-plasma conditions, a high-density electron layer can be formed by an intense driver laser pulse in the plasma medium, and it propagates or oscillates with a relativistic speed. The high-density electron layer behaves as a mirror and a counter-propagating laser pulse (known as the source laser pulse) to the electron layer can be reflected and focused to a high intensity. Since the reflected laser pulse experiences the double Doppler effect, the frequency of the reflected laser is upshifted by $4 \gamma^2$, where $\gamma$ is the Lorentz $\gamma$-factor for the high-density electron layer. Interestingly, the field strength of the reflected laser pulse is enhanced by a factor of $\gamma^3 D/\lambda$ \cite{Bulanov1994, Jeong_RFM}, where $D$ is the beam diameter of the driver laser pulse. This implies that the required laser intensity for reaching the critical field strength is reduced to
%%%% Equation 1-4
\begin{equation} \label{eq1-4}
I = \frac{\left( 2\lambda / D \right)^2}{6 \pi^5 \gamma^6 \mathcal{R}} I_c,
\end{equation}
where $\mathcal{R}$ is the reflectance of the mirror and $I_c$ the critical intensity. However, there are two points to be considered in the practical application of the relativistic flying mirror. One is the reflectance of the relativistic flying mirror and the other is the peak power of source laser. The reflectance of the relativistic mirror is in general very low, and it depends on the electron density and the $\gamma$-factor of the mirror \cite{Kulagin, Esirkepov, Bulanov_AM}. The source laser power is related to the recoil effect \cite{Valenta}, so the source laser power should be sufficiently low ($\eta \ll 1$) not to seriously affect the relativistic flying mirror. Recent investigation shows that despite the low reflectance and the low source laser power the laser intensity focused by the relativistic flying mirror can surpass the conventional laser intensity obtained by the optical-focusing scheme as shown in this paper.

In this review, we discuss how to obtain a super-strong field with a relativistic flying mirror and how to apply such a super-strong field to fundamental science, specifically to the SF-QED in the non-perturbative regime. The paper is organized as follows. In section 2, the characteristics of a fs high-power laser pulse and its focusing schemes, including the normal-focusing, tight-focusing, and 4$\pi$-spherical-focusing schemes, are briefly reviewed for the estimation of attainable laser intensity through the optical approach. In section 3, the relativistic flying mirror has been discussed. The reflection coefficient and the field strength of the focused source laser pulse are discussed. In section 4, it is shown that the field strength obtained by the ideal relativistic flying mirror can be sufficiently high enough for the study on the SF-QED in the non-perturbative regime.

\section{High-power laser pulse and attainable laser intensity}

\subsection{Brief overview of the high-power laser pulse}
The fs PW-class high-power laser based on the CPA technique consists of the fs laser oscillator, the pulse stretcher, amplifier chains, and the pulse compressor \cite{Jeong_Laser}. A femtosecond laser oscillator generates a transform-limited laser pulse. The fs Ti:sapphire laser that typically produces 7-8 fs laser pulses is commonly used as a laser oscillator in the front-end. The fs laser pulse is temporally stretched to $\sim$ns level before the energy amplification. The stretched laser pulse experiences the gain narrowing effect during the amplification, so the fs high-power laser pulse has a longer pulse duration (20 fs - 40 fs) after compression. The contrast of the fs high-power laser pulse plays an important role in the laser-matter interaction \cite{Jeong_RHHG,kiriyama}, but the contrast is not discussed in this paper.  

After compression, a fs high-power laser pulse can be in general described by
%%% Equation 2-1-1
\begin{equation} \label{eq2-1-1}
E_i(x,y,z;t) = E_0 \cdot \mathcal{F} (x,y) \cdot \exp [ ik \mathcal{W} (x,y) ] \cdot \mathcal{T} (z;t).  
\end{equation}
Here, $\mathcal{F} (x,y)$ is the spatial distribution function of the electric field. For the high-power laser pulse, the laser beam has a flat-top beam profile to ensure high-energy-extraction efficiency and to reduce the optical damage induced by the nonlinear effect. Thus, it is common to express the spatial distribution function as a super-Gaussian (SG) beam to represent a flat-top (uniform) beam profile. Recently, the Laguerre-Gaussian (LG) beam profile \cite{Zauderer} is widely considered for specific applications \cite{Jentschura,Dumlu}. The mathematical forms for the SG and the LG beams are
%%% Equation 2-1-2
\begin{equation} \label{eq2-1-2}
\mathcal{F} (x,y) = 
\begin{cases} 
\exp \left( -\frac{r^{2n}}{w_0^2} \right) \\
\left( \frac{\sqrt{2}r}{w_0} \right)^{|l|} L_n^l \left( - \frac{2 r^2}{w_0^2} \right) \exp \left( -\frac{r^2}{w_0^2} \right) \exp (-il \phi)
\end{cases},
\end{equation}
where $L_n^l (\cdot)$ is the associated Laguerre polynomial, $w_0$ the Gaussian width, and $l$ the topological charge, respectively. 

The wavefront aberration, $\mathcal{W} (x,y)$, is an optical error induced by the imperfection of optical elements and the thermal property of the amplification medium. The high-power laser pulse has wavefront aberration and the focused laser spot is mostly determined by the wavefront aberration. The wavefront aberration is commonly represented by the Zernike polynomials or the Seidel aberration modes \cite{Malacara}. The wavefront aberration should be corrected by an adaptive optics (AO) system before being focused on a target \cite{Jeong_AO}. The Strehl ratio or residual wavefront aberration is used to assess the focal spot quality and the focused intensity on target \cite{Pirozhkov}. A more sophisticated method based on the phase retrieval algorithm can provide the focused field information at different wavelengths \cite{Borot}. Despite the residual wavefront aberration plays a critical role for the practical laser intensity, we assume in this paper that the wavefront aberration is perfectly corrected by an AO system. 

The temporal distribution function, $\mathcal{T} (z;t)$, is related to the temporal pulse profile. The temporal distribution function is given by
%%%% Equation 2-1-3
\begin{equation} \label{eq2-1-3}
\mathcal{T} (z;t) = e^{ikz} \int_{-\infty}^{\infty} s(\omega) e^{i \varphi (\omega)} e^{-i \omega t} d \omega,
\end{equation} 
where $|s(\omega)|^2$ is the spectral intensity (commonly the laser spectrum) and $\varphi (\omega)$ is the spectral phase. The spectral phase plays a crucial role in determining the laser pulse temporal shape. The spectral phase is expressed as \cite{Diels},
%%%% Equation 2-1-4
\begin{equation} \label{eq2-1-4}
\varphi (\omega) = \sum_{n=0}^{\infty} \frac{\beta_n}{n!} (\omega - \omega_c)^n,
\end{equation} 
with the definition of $\beta_n = \frac{d^n \varphi (\omega)}{d \omega^n} \big|_{\omega = \omega_c}$. The second derivative term, $\beta_2$, is known as the group delay dispersion (GDD), which is responsible for the linear frequency-chirping of an ultrashort laser pulse and results in the pulse broadening in time. Higher-order terms, such as $\beta_3$, $\beta_4$, and so on, are known as third-order dispersion (TOD), fourth-order dispersion (FOD), and so on. These terms are responsible for the non-linear frequency chirping and the deformation of the laser temporal shape. The spectral phase can be pre-compensated by using an acousto-optic programmable dispersive filter (AOPDF) \cite{Tournois}. Recently, it is known that focusing a chirped laser pulse with a dispersive lens can produce a super-luminal or sub-luminal laser pulse near focus \cite{Froula,Jeong_Super}. 

The fs high-power laser pulse usually contains a small amount of wavefront aberration [Fig. 1(a)] and spectral phase [Fig. 1(c)]. Figure 1(b) and 1(d) show the practical focal spot and temporal profile of the fs high-power laser pulse with those errors. The wavefront aberration and the spectral phase deviate the spatiotemporal profile from the ideal case. However, we consider the ideal case in this paper. Thus, we regard $\exp [ ik \mathcal{W} (x,y) ] = 1$ and $\mathcal{T} (z;t) = e^{ikz} \int_{-\infty}^{\infty} s(\omega) e^{-i \omega t} d \omega$ assuming $\mathcal{W} (x,y) = 0$ and $\varphi (\omega) =0$.

%%%%%%%%%%% Figure 1
\begin{figure}%[b]
\centering
\includegraphics[width=1\columnwidth]{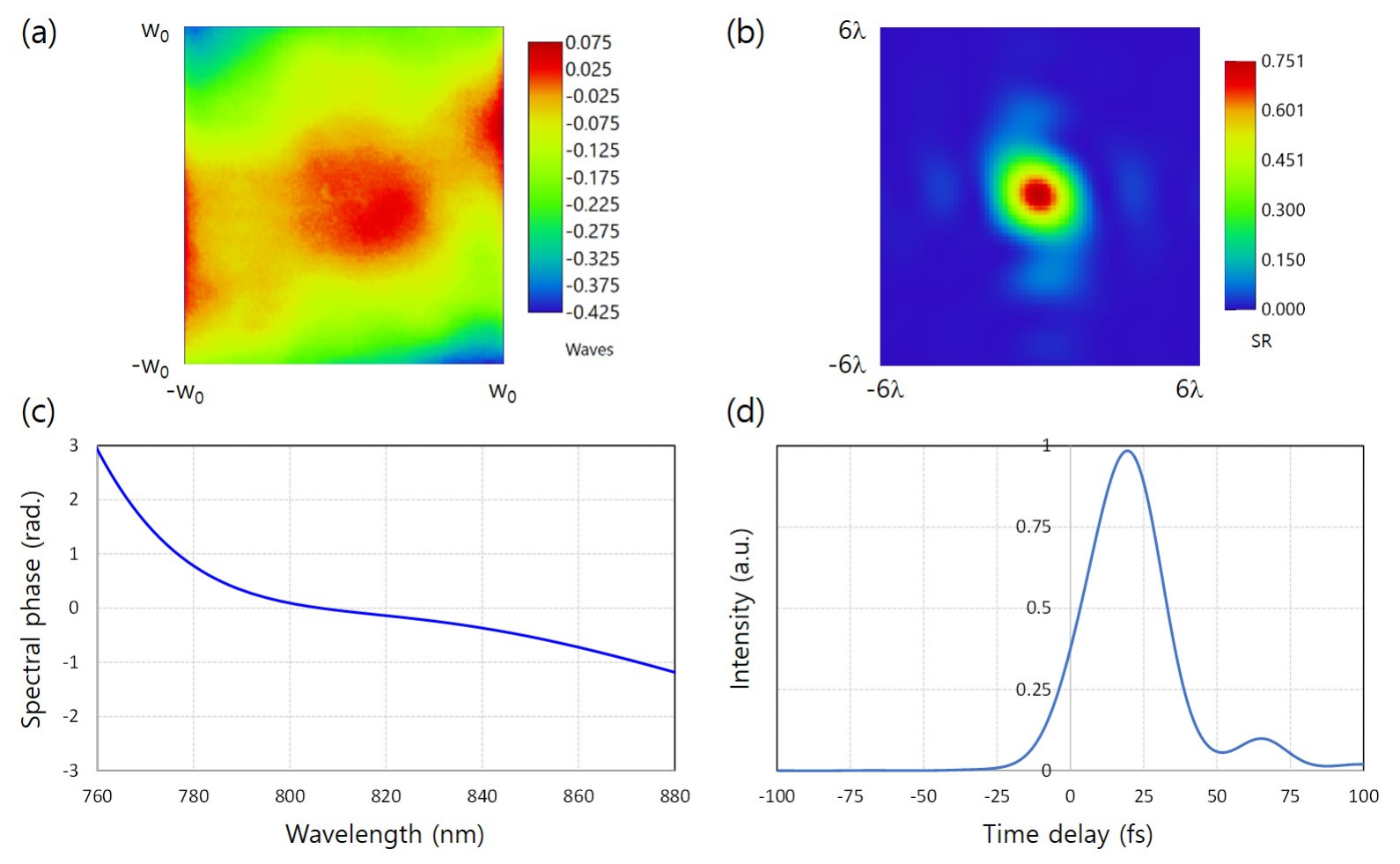}% Here is how to import EPS art \caption{\label{fig:epsart}
\caption{Effect of the wavefront aberration (a) and the spectral phase (c) on the focal spot (b) and the temporal profile (d). The wavefront aberration is a resultant aberration accumulating all wavefront in a beam transport line in a high-power laser system. The spectral phase is measured in the vacuum for the p-polarization. A typical high-power Ti:Sapphire laser spectrum to support a pulse duration of 30 fs is used. The GDD is compensated in the calculation.}
\end{figure}

\subsection{Normal focusing scheme of the laser beam: paraxial approximation}

The laser pulse is focused by an optic with a focal length of $f$ to reach a required field strength. Under the normal focusing scheme, the paraxial approximation is used, and the focused laser field, $E_f$, is described by the Huygens-Fresnel diffraction integral as
%%%% Equation 2-2-1
\begin{equation} \label{eq2-2-1}
d E_f = \frac{i \omega E_i}{2 \pi c} e^{-i \frac{k \rho^2}{2f}} \frac{e^{i \vec{k} \cdot (\vec{r} - \vec{\rho})}}{|\vec{r} - \vec{\rho}|} dS,
\end{equation}
where $\vec{k}$ is the wavevector of the laser pulse, $\vec{r}$ the vector from the origin to an observation position, $\vec{\rho}$ the vector from the origin to a source position, and $dS$ the infinitesimal area on the source plane. A uniform field strength, $E_i$, is assumed within a beam radius of $r_0$. The quadratic phase, $k \rho^2 / 2 f$, is introduced by the focusing optic. In carrying out the integration for Eq. \eqref{eq2-2-1}, a common technique is to replace $\vec{k} \cdot (\vec{r} - \vec{\rho})$ and $|\vec{r} - \vec{\rho}|$ with
%%%% Equation 2-2-2
\begin{equation} \label{eq2-2-2}
\vec{k} \cdot (\vec{r} - \vec{\rho}) \approx k (z + f) - \frac{kr \rho \cos (\phi - \varphi) }{z+f} + \frac{k (r^2 + \rho^2 )}{2(z+f)}, \quad \textrm{and} \quad |\vec{r} - \vec{\rho}| \approx f.
\end{equation}
The $\phi$ and $\varphi$ in Eq. \eqref{eq2-2-2} stand for the azimuthal angles in the observation and the source planes. By inserting Eq. \eqref{eq2-2-2} into Eq. \eqref{eq2-2-1}, we obtain
%%%% Equation 2-2-3
\begin{equation} \label{eq2-2-3}
d E_f \approx i \frac{\omega f E_i}{2 \pi c} e^{i k (z+f)} \exp \left( i \frac{k r^2}{2z} \right) \exp \left[ - i \frac{k r \rho \cos (\phi -\varphi)}{f} \right] \exp \left(  i \frac{k z}{2f^2} \rho^2 \right) \sin \theta d \theta d \varphi. 
\end{equation}
Under the paraxial approximation, $\sin \theta$ is replaced by $\theta$ and $\rho$ by $f \theta$. The Jacobi-Anger identity given by
%%%% Equation 2-2-4
\begin{equation} \label{eq2-2-4}
\int_0^{2 \pi} e^{i l \phi} e^{i x \cos (\phi - \phi') } d \phi = 2 \pi i^l e^{il \phi'} J_l (x)
\end{equation}
yields 
%%%% Equation 2-2-5
\begin{equation} \label{eq2-2-5}
d E_f \approx i \frac{\omega f E_i}{c} e^{i k (z+f)} \exp \left( i \frac{k r^2}{2f} \right) J_0 (k r \theta ) \exp \left( i \frac{k}{2} z \theta^2 \right) \theta d \theta.
\end{equation}
Now, by separating Eq. \eqref{eq2-2-5} into two integrals defined as,
%%%% Equation 2-2-6
\begin{equation} \label{eq2-2-6}
\mathcal{I}_1 = \int_0^{\theta_0} J_0 (kr \theta) \cos \left( \frac{k}{2} z \theta^2 \right) \theta d \theta, \quad \textrm{and} \quad \mathcal{I}_2 = \int_0^{\theta_0} J_0 (kr \theta) \sin \left( \frac{k}{2} z \theta^2 \right) \theta d \theta,
\end{equation}
and carrying out the integration with the Bessel function identity given as,
%%%% Equation 2-2-7
\begin{equation} \label{eq2-2-7}
\frac{d}{dx} \left[ x^{n+1} J_{n+1} (x) \right] = x^{n+1} J_n (x),
\end{equation}
we obtain the focused field expression near focus as,
%%%% Equation 2-2-8
\begin{equation} \label{eq2-2-8}
E_f \approx i \frac{\omega f E_i}{c} \theta_0^2 e^{i \left[ k (z+f) +  \frac{k r^2}{2f} +  \frac{kz}{2} \theta_0^2 \right]} \left[ \frac{J_1 \left( k r \theta_0 \right)}{k r \theta_0} - ik z \theta_0^2 \frac{J_2 \left( k r \theta_0 \right)}{ \left( k r \theta_0 \right)^2} \right] .
\end{equation}
Here, $\theta_0$ is $\tan^{-1} (r_0 / f) \approx r_0 /f$. Under the paraxial approximation ($\theta_0 \ll 1$), higher-order terms of $kz \theta_0^2$ are neglected. In the focal plane ($z=0$), the first zero of $J_1 (k r \theta_0)$ exists at $r \approx 1.22 \lambda f / 2r_0$, or $r  \approx 1.22 \lambda f/\#$ with the definition of the f-number ($f/\# \equiv f/ 2 r_0$). A practical definition for the size of focused field is the full width at the half maximum (FWHM). For the Ti:Sapphire laser, the FWHM for the uniform input electric field is approximately given by $\lambda f/\#$. 

As the f-number decreases below 1, which is known as the tight focusing condition, the above approach based on the scalar diffraction theory is not valid in describing the field distribution and in predicting the peak intensity of the focused laser pulse (See Fig. 2). The vector diffraction theory should be applied to this case. 

%%%%%%%%%%% Figure 2
\begin{figure}%[b]
\centering
\includegraphics[width=0.6\columnwidth]{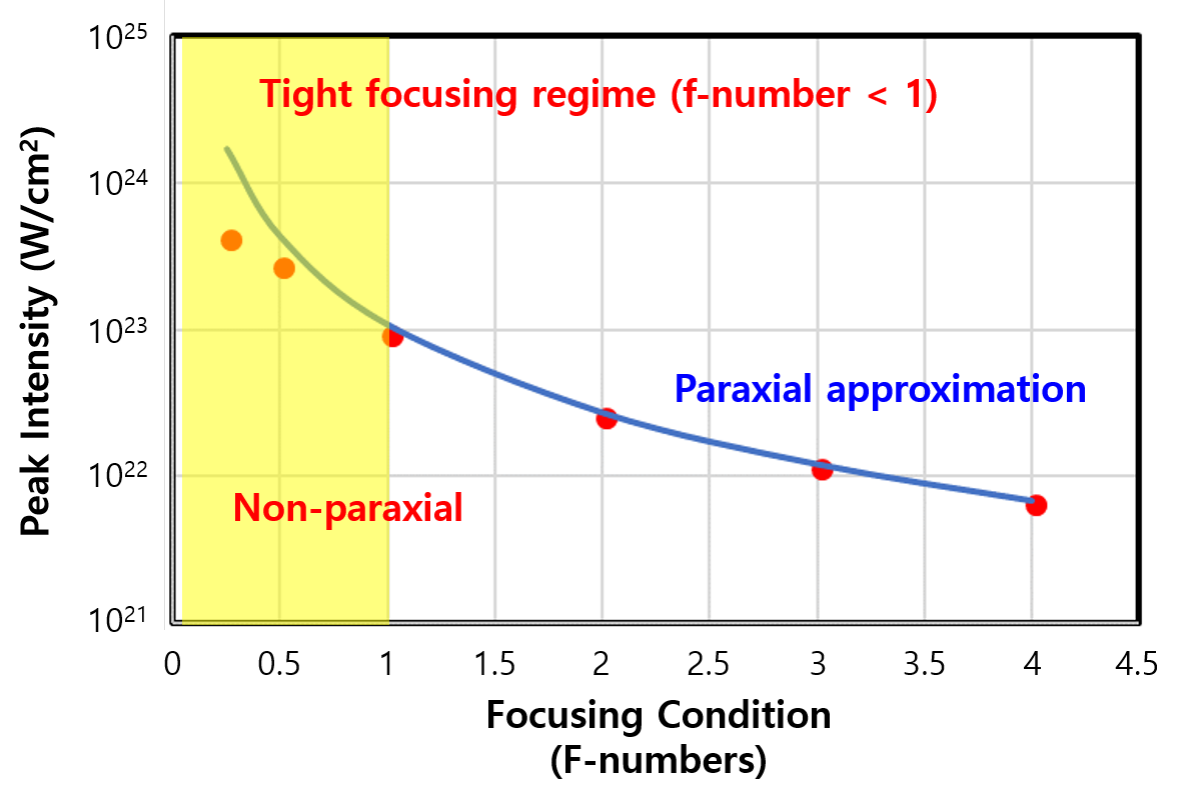}% Here is how to import EPS art \caption{\label{fig:epsart}
\caption{The theoretical peak intensity obtained with a 1 PW laser pulse with respect to the f-number. The blue line is calculated based on the geometrical optics. The red circles are calculated from the wave optics based on the vector differential integral equation [Eq. \eqref{eq2-3-1}]. Both approaches yield the same peak intensity in the paraxial region where f-number $\gg$1. The intensities start to deviate from each other under the condition of f-number <1.}
\end{figure}

\subsection{Tight focusing and 4\texorpdfstring{$\pi$-}{pi-}spherical focusing schemes}
Under the tight focusing condition, the incident laser pulse is focused by a low f-number OAP mirror. At an observation point ($x_0$, $y_0$, $z_0$), the field reflected and focused by a paraboloidal surface can be calculated by the Sttraton-Chu's diffraction integral given as \cite{Stratton,Varga},
%%%% Equation 2-3-1
\begin{equation} \label{eq2-3-1}
E_{f,j} = i \frac{\omega f E_i}{\pi c} e^{i 2 k f} \int_0^{2 \pi} \int_{\theta_{max}}^{\pi} \frac{\alpha_j (\theta_s , \phi_s)}{1 - \cos \theta_s} \exp [ - i k \Phi (x_o, y_o, z_o, \theta_s, \phi_s) ] d \Omega_s. 
\end{equation}
Here, the subscript, $j$, stands for $x$-, $y$-, or $z$-polarization states, and the subscript, $s$, is used to denote a source position on the parabolic surface. The phase factor, $\Phi (x_o, y_o, z_o, \theta_s, \phi_s)$, is given by
%%%% Equation 2-3-2
\begin{equation} \label{eq2-3-2}
\Phi (x_o, y_o, z_o, \theta_s, \phi_s) = z_o \cos \theta_s + x_o \sin \theta_s \cos \varphi_s + y_o \sin \theta_s \sin \varphi_s .
\end{equation}
The $d \Omega_s$ is the infinitesimal solid angle at the source position ($\theta_s$,$\phi_s$) on the mirror, where $\theta_s$ and $\phi_s$ are the polar and the azimuthal angles to a source point on the mirror. 

The expression, $\alpha_j$, in the integrand of Eq. \eqref{eq2-3-1} depends on the polarization of the input laser pulse. For the linearly-polarized input field, the $\alpha_j$s are given by \cite{Jeong_LTF},
%%%% Equation 2-3-3
\begin{eqnarray} \label{eq2-3-3}
\alpha_x &=& 1- \frac{\sin \theta_s \cos \phi_s}{1 - \cos \theta_s} \left( 1 - \frac{1- \cos \theta_s}{i 2 k f} \right) [ 2 f \sin \theta_s \cos \phi_s - x_o (1 - \cos \theta_s)], \\
\alpha_y &=& \frac{\sin \theta_s \cos \phi_s}{1 - \cos \theta_s} \left( 1 - \frac{1- \cos \theta_s}{i 2 k f} \right) [ 2 f \sin \theta_s \sin \phi_s - y_o (1 - \cos \theta_s)], \\
\textrm{and} \nonumber \\
\alpha_z &=& \frac{\sin \theta_s \cos \phi_s}{1 - \cos \theta_s} \left\{ 1 - \left( 1 - \frac{1- \cos \theta_s}{i 2 k f} \right) [ 2 f \cos \theta_s - z_o (1 - \cos \theta_s)] \right\}.
\end{eqnarray}
For the radially-polarized input laser, the $\alpha_j$s are given by \cite{Jeong_RATF}
%%%% Equation 2-3-4
\begin{eqnarray} \label{eq2-3-4}
\alpha_x &=& \cos \phi_s - \frac{\sin \theta_s }{1 - \cos \theta_s} \left( 1 - \frac{1- \cos \theta_s}{i 2 k f} \right) [ 2 f \sin \theta_s \cos \phi_s - x_p (1 - \cos \theta_s)], \\
\alpha_y &=& \sin \phi_s - \frac{\sin \theta_s }{1 - \cos \theta_s} \left( 1 - \frac{1- \cos \theta_s}{i 2 k f} \right) [ 2 f \sin \theta_s \sin \phi_s - y_p (1 - \cos \theta_s)], \\
\textrm{and} \nonumber \\
\alpha_z &=& \frac{\sin \theta_s }{1 - \cos \theta_s} \left\{ 1 - \left( 1 - \frac{1- \cos \theta_s}{i 2 k f} \right) [ 2 f \cos \theta_s - z_p (1 - \cos \theta_s)] \right\}.
\end{eqnarray}
For the azimuthally-polarized input laser, the $\alpha_j$s are given by \cite{Jeong_RATF}
%%%% Equation 2-3-5
\begin{eqnarray} \label{eq2-3-5}
\alpha_x &=& - \sin \phi_s, \\
\alpha_y &=& \cos \phi_s, \\
\textrm{and} \nonumber \\
\alpha_z &=& 0.
\end{eqnarray}
The focused field contains three (two transverse and one longitudinal) polarization components. 
For linear and radial polarizations, the longitudinal field can be the dominant component in the intensity distribution as the f-number decreases (See Fig. 3). The calculation shows that the highest strength of 2.2$\times$10$^{13}$ V/cm can be reached for the longitudinal field by tightly focusing a 1 PW radially-polarized laser pulse.

%%%%%%%%%%% Figure 3
\begin{figure}%[b]
\centering
\includegraphics[width=1\columnwidth]{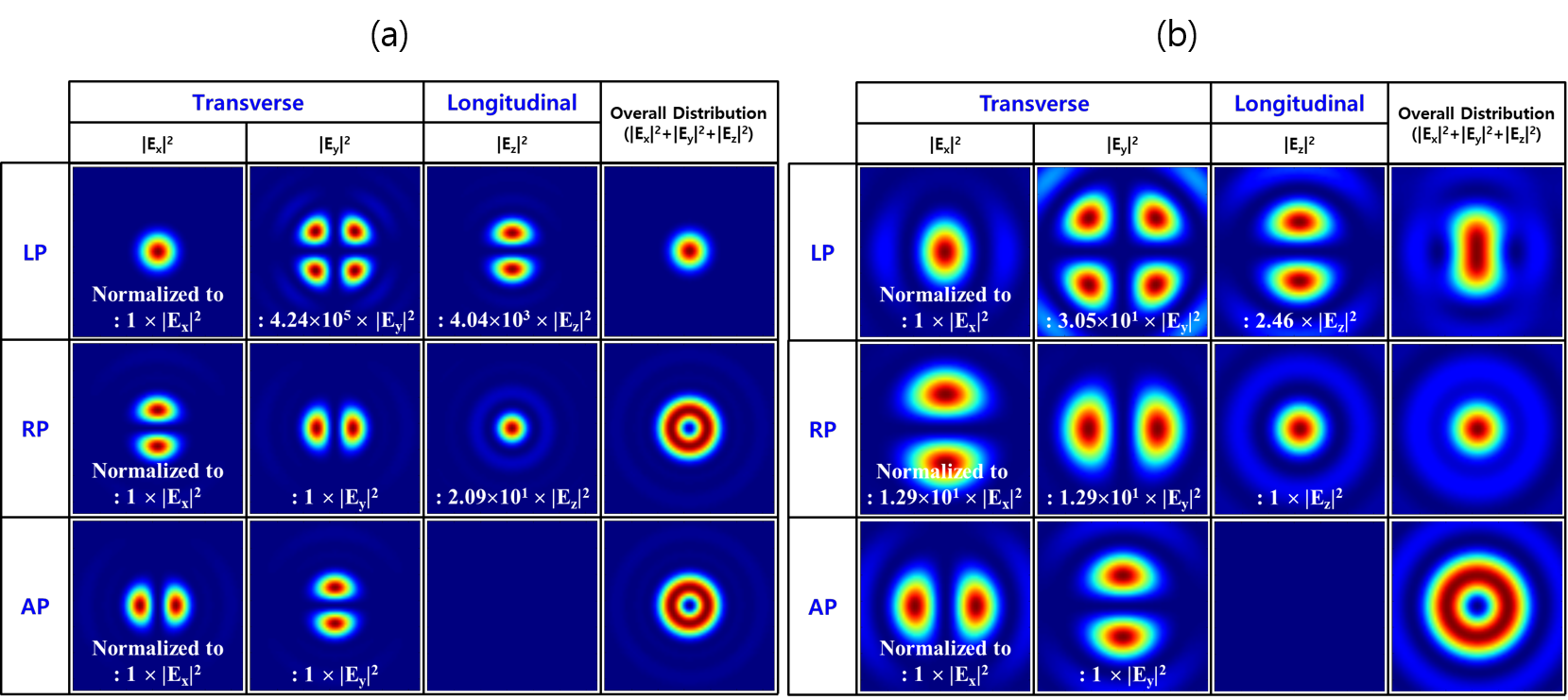}% Here is how to import EPS art \caption{\label{fig:epsart}
\caption{Intensity distributions obtained with (a) the normal focusing condition (f/\# = 3) (b) the tight focusing condition (f/\# = 0.25). Three different input polarization states, such as LP (linear polarization along the x-direction), RP (radial polarization), and AP (azimuthal polarization), are considered. The input laser pulse is x-polarized. The $|E_x|^2$ and $|E_y|^2$ stand for the transverse components and the $|E_z|^2$ does the longitudinal (or axial) component. Modified from Figs. 2 and 3 in \cite{Jeong_RATF}.}
\end{figure}

The 4$\pi$-spherical focusing scheme is regarded as the extreme case of either tight focusing (in which the f-number approaches zero) or multiple laser beam focusing (in which
the number of beams approaches infinity) scheme \cite{SSBulanov,Jeong_4pi}. For the TM mode (radially-polarized input laser beam) case, the electric and magnetic field distributions obtained by the 4$\pi$-spherical focusing scheme are given by,
%%%% Equation 2-3-6
\begin{equation} \label{eq2-3-6}
\vec{E}_f = \hat{\theta} i \frac{\pi}{2} k r_s E_s e^{ikr_s} a(r, \theta), \quad \textrm{and} \quad \vec{B}_f = - \hat{\phi} \frac{\pi}{2} k r_s \frac{E_s}{c} e^{ik r_s} b (r, \theta). 
\end{equation}
Here, $k r_s$ is the phase from a focus point to a virtual sphere and it can be ignored without any problem. The field strength on the virtual surface, $E_s$, is given by $\sqrt{3 P_L / 4 \pi c \varepsilon_0 r_s^2}$ with an incident laser power, $P_L$. In Eq. \eqref{eq2-3-6}, $a (r, \theta)$ and $b (r, \theta)$ are given by,
%%%% Equation 2-3-7
\begin{eqnarray} \label{eq2-3-7}
a (r, \theta) &=& j_0 (kr) + \frac{5}{2^3} j_2 (kr) P_2 ( \cos \theta) - \frac{9}{2^6} j_4 (kr) P_4 (\cos \theta) + \cdots , \\
\textrm{and} \nonumber \\
b (r, \theta) &=& \frac{4}{\pi} j_1 (kr) P_1^1 ( \cos \theta), 
\end{eqnarray}
where $j_n (\cdot)$ is the spherical Bessel function, $P_n (\cdot)$ the Legendre polynomial, and $P_n^m (\cdot)$ the associated Legendre polynomial, respectively. Figure 4 shows the focused electric and magnetic field configurations by the 4$\pi$-spherical focusing scheme for the TM mode. For the azimuthally-polarized input laser beam (TE mode), the electric and magnetic field distributions by the 4$\pi$-spherical focusing scheme are given by
%%%% Equation 2-3-8
\begin{equation} \label{eq2-3-8}
\vec{E}_f = \hat{\phi} i \frac{\pi}{2} k r_s E_s e^{ikr_s} b(r, \theta), \quad \textrm{and} \quad \vec{B}_f = - \hat{\theta} \frac{\pi}{2} k r_s \frac{E_s}{c} e^{ik r_s} a (r, \theta), 
\end{equation}
due to the polarization symmetry. The calculation based on the dipole radiation \cite{Gonoskov_4pi} shows a similar field distribution. According to \cite{SSBulanov}, the 4$\pi$-spherical focusing scheme with the linear polarization shows multiple intensity peaks. However, the radial and the azimuthal polarization provide smoother intensity distribution.

%%%%%%%%%%% Figure 4
\begin{figure}%[b]
\centering
\includegraphics[width=1\columnwidth]{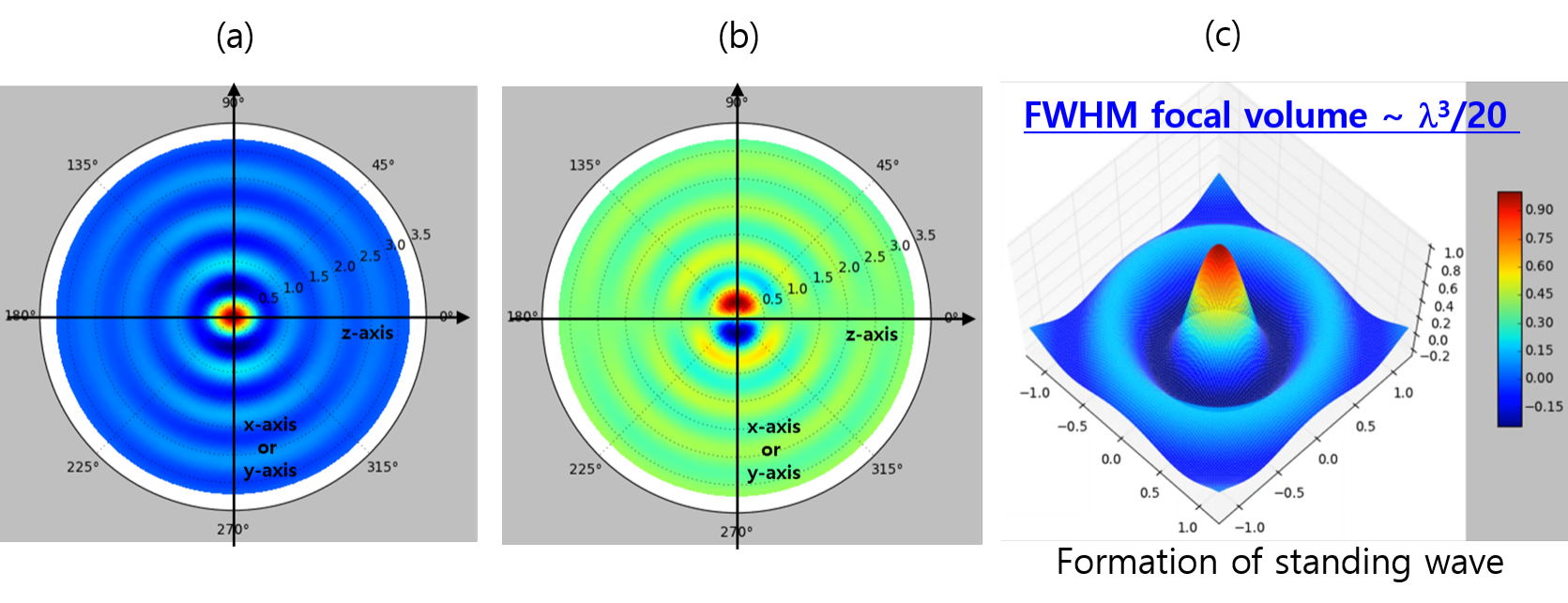}% Here is how to import EPS art \caption{\label{fig:epsart}
\caption{(a) Focused electric field by the 4$\pi$-spherical focusing scheme, (b) Focused magnetic field by the 4$\pi$-spherical focusing scheme, (c) Electric field in 3D format. The input laser pulse is radially polarized (TM mode). For the azimuthally polarized (TE mode) input laser pulse, the electric (a) and magnetic (b) fields are switched.}
\end{figure}

The highest electric field strength can be obtained with the TM mode EM wave at the focus as,
%%%% Equation 2-3-9
\begin{equation} \label{eq2-3-9}
E_p = \frac{k}{4} \sqrt{\frac{3 \pi P_L}{\varepsilon_0 c}}. 
\end{equation}
And, the focal volume, $V_f$, for this case is approximated as 
%%%% Equation 2-3-10
\begin{equation} \label{eq2-3-10}
V_f \approx \frac{\lambda^3}{20}. 
\end{equation}
The 4$\pi$-spherical focusing scheme provides the highest field strength attainable with a given laser power through the optical-focusing scheme. For example, the highest laser field strength with a 10 PW Ti:S laser pulse ($\lambda = 0.8 \mu$m) is about $1.17 \times 10^{14}$ V/cm, which is about one-hundredth of the critical field strength. 

\section{Super-strong field formation through the relativistic flying parabolic mirror}

\subsection{Relativistic flying parabolic mirror}

We assume that the relativistic flying mirror has a parabolic curved surface \cite{Matlis,Wan}. The focused laser beam profile can be described by either the Gaussian or Bessel function ($J_1 ( k r \theta_0) / k r \theta_0$ as shown in Eq. \eqref{eq2-2-8}). Under the first-order approximation, both functions can be expressed by a quadratic function ($1 - k^2 r^2 \theta_0^2$). The shape of a relativistic flying mirror can be approximated by the paraboloidal shape under the first-order approximation. This mirror is regarded as the relativistic flying parabolic mirror (RFPM). It should be noted that the formation of a relativistic spherical plasma wave is also studied in \cite{SSBulanov_PoP}.

Due to the special relativity, the focal length of the RFPM in the boosted frame of reference can be much shorter than the diameter of the RFPM (See Fig. 5). This can be justified in the following approach. In this case, the field expressions obtained by the 4$\pi$-spherical focusing scheme can be used to describe the field configuration by the RFPM. In the boosted frame of reference (primed coordinate system), the equation for the paraboloidal surface is given by
%%%% Equation 3-1-1
\begin{equation} \label{eq3-1-1}
z' = \frac{{x'}^2 + {y'}^2}{4 f'} -f'. 
\end{equation}
where $f'$ is the focal length of the mirror in the boosted frame. Since the RFPM moves along the +$z$ axis, the Lorentz transformations between the laboratory (unprimed coordinate system) and the boost frames are given as
%%%% Equation 3-1-2
\begin{equation} \label{eq3-1-2}
x' = x, \quad y' = y, \quad z' = \gamma (z - \beta c t), \quad \textrm{and} \quad ct' = \gamma ( ct - \beta z). 
\end{equation}
By using the above Lorentz transformation, the equation for the paraboloidal surface in the laboratory frame can be expressed as,
%%%% Equation 3-1-3
\begin{equation} \label{eq3-1-3}
z = \frac{x^2 + y^2}{4 \gamma f'} - \gamma f' + \frac{\gamma^2 - 1}{\gamma} f' + \beta c t. 
\end{equation}
From Eq. \eqref{eq3-1-3}, it is obvious that the focal length in the laboratory frame is modified as $\gamma f'$. This means that a focal length observed in the laboratory frame is $\gamma$-times longer than that in the boosted frame. For example, a focal length of 10 to a few tens of $\mu$ms in the laboratory frame is shortened by less than a few $\mu$ms in the boosted frame with a $\gamma$-factor of 20, yielding a very low $f/\#$ in the boosted frame with a typical diameter of a few tens of $\mu$ms for the relativistic mirror. Thus, the 4$\pi$-spherical focusing scheme can be applied to calculate the field distribution in the boosted frame. 

%%%%%%%%%%% Figure 5
\begin{figure}[b]
\centering
\includegraphics[width=0.9\columnwidth]{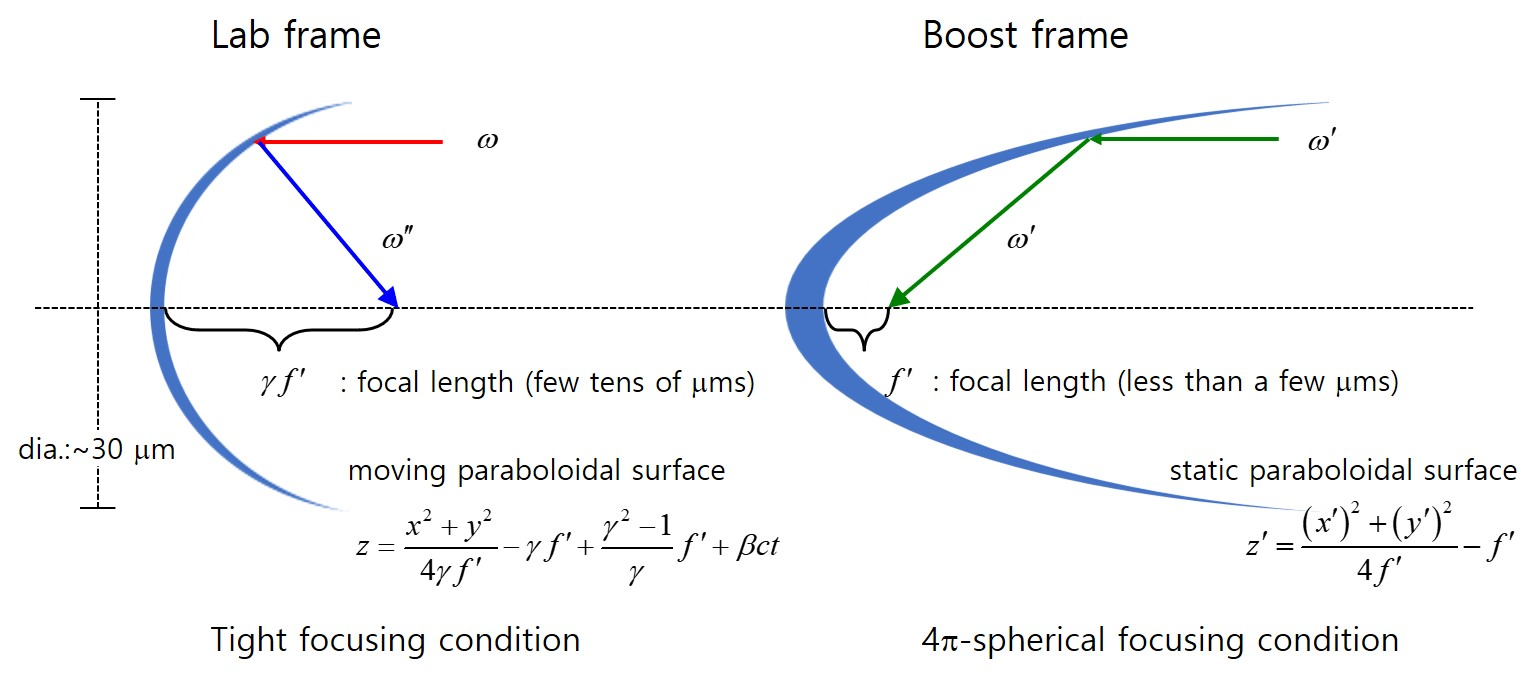}% Here is how to import EPS art \caption{\label{fig:epsart}
\caption{Relativistic flying parabolic mirror in the laboratory and the boosted frame of reference. Due to the short focal length compared to the mirror size, the incident laser pulse is 4$\pi$-spherically focused in the boosted frame.}
\end{figure}

Now, let us consider the recoil effect occurring in the relativistic flying mirror. The recoil effect can be examined through the energy-momentum conservation as follows. By assuming the rotational symmetry, we may write the energy-momentum conservation in the two-dimensional form as \cite{Valenta},
%%%% Equation 3-1-4
\begin{equation} \label{eq3-1-4}
n_e p_e - n_{\omega} p_{\omega} = n_e p''_e \cos \theta_e + \mathcal{R} n_{\omega} p''_{\omega} \cos \theta - ( 1 - \mathcal{R}) n_{\omega} p_{\omega} ,
\end{equation}
%%%% Equation 3-1-5
\begin{equation} \label{eq3-1-5}
0 = n_e p''_e \sin \theta_e - \mathcal{R} n_{\omega} p''_{\omega} \sin \theta,
\end{equation}
and
%%%% Equation 3-1-6
\begin{equation} \label{eq3-1-6}
n_e \varepsilon_e + n_{\omega} \varepsilon_{\omega} = n_e \varepsilon''_e + \mathcal{R} n_{\omega} \varepsilon''_{\omega} + ( 1 - \mathcal{R} ) n_{\omega} \varepsilon_{\omega}.
\end{equation}
Here, $p$ and $\varepsilon$ refer to the momentum and energy for the individual electron and photon, and the subscripts $e$ and $\omega$ are used to denote the electron and photon. The population densities for the electron and photon are denoted by $n_e$ and $n_{\omega}$, respectively. The $\mathcal{R}$ again stands for the reflectance of the relativistic flying mirror. The unprimed and double-primed quantities refer to the quantities before and after reflection in the laboratory frame. The angles, $\theta_e$ and $\theta$, refer to the angles for the electron and photon after reflection. 

The energy densities for electrons and photons may be rewritten as,
%%%% Equation 3-1-7
\begin{equation} \label{eq3-1-7}
n_e \varepsilon_e = \gamma n_e m c^2, \quad \textrm{and} \quad n_{\omega} \varepsilon_{\omega} = I_s/c.
\end{equation}
Here, $I_s$ is the intensity of the source laser pulse. From the energy-momentum relation, we write $\varepsilon_e^2 - p_e^2 c^2 = {\varepsilon''_e}^2 - {p''_e}^2 c^2 = m^2 c^4$ for electron and $\varepsilon_{\omega}^2 - p_{\omega}^2 c^2 = {\varepsilon''_{\omega}}^2 - {p''_{\omega}}^2 c^2 = 0$ for photon. By plugging above relationships into the energy-momentum conservation law, we obtain the angular frequency for the reflected EM wave given as,
%%%% Equation 3-1-8
\begin{equation} \label{eq4-1-10}
\omega'' \approx \omega \frac{1 + \beta}{1 - \beta \cos \theta} \left[ 1 - \frac{\mathcal{R} I_s}{\gamma n_e m c^3} (1 + \cos \theta )^2 \right] .
\end{equation}
The second term containing $\mathcal{R} I_s / \gamma n_e m c^3$ in the parenthesis on the right-hand side is responsible for the recoil effect. The second term can be ignored when the reflected energy density ($\mathcal{R}I_s/c$) is sufficiently lower than the energy density of the electrons in the RFPM ($\gamma n_e m c^2$). This condition is satisfied by a counter-propagating source laser pulse with a low $\eta$ ($\ll 1$).

\subsection{Reflection coefficient of the relativistic flying mirror}
In the laboratory frame, a source laser pulse ($\eta < 1$) collides with energetic electrons ($\gamma \gg 1$), so the incident radiation is backward-scattered (radiated) from the electrons since $\gamma (1 + \beta) \gg 1 + \eta^2$ \cite{Esarey,Ehlotzky}. Since the source laser is weak, the oscillation speed of electrons in the boosted frame is much smaller than the speed of light, and the scattered radiation can be treated as coherent radiation. This is easily verified by comparing the photon formation length to the Compton wavelength, $\lambdabar$, defined as $\lambdabar = \hbar/mc$. The photon formation length shorter than the Compton wavelength makes radiated photons incoherent since the radiation process occurs in a distinct region and its phase becomes stochastically random. On the other hand, the photon formation length longer than the Compton wavelength makes photons coherent. The photon formation length, $l_f$, is defined as
%%%% Equation 3-2-1
\begin{equation} \label{eq3-2-1}
l_f = 2 r_c \theta, 
\end{equation}
where $\theta$ is related to the emission angle through $1/\gamma$ and $r_c$ is the instantaneous radius of curvature for the electron trajectory given by \cite{Gonoskov_RMP}
%%%% Equation 3-2-2
\begin{equation} \label{eq3-2-2}
r_c = \frac{\lambdabar (\gamma^2 -1 )}{\sqrt{\chi_e^2 - (\vec{E} + \vec{\beta} \times \vec{B} )^2 / E_c^2}}. 
\end{equation}
Here, $\vec{E}$ and $\vec{B}$ are the electric and magnetic fields of the source laser incident on the relativistic flying mirror. When $\beta \ll 1$, the instantaneous radius can be approximated as $\lambdabar \eta E_c /E_i$ and the photon formation length is much longer than the Compton wavelength. Thus, electromagnetic radiation reflected from the relativistic flying mirror can be treated as classical coherent radiation. 

When a relativistic flying mirror driven by an intense EM wave (driver wave) encounters a weakly relativistic EM wave (source wave), the general form for the reflectivity of a relativistic flying mirror may be calculated through the Lienard-Wiechert field as follows. We treat the relativistic flying mirror as a collisionless plasma traveling along the +$z$-direction. Assuming that the incident wave with an angular frequency, $\omega$, is x-polarized [$\vec{A} = A e^{i \omega t + \phi (z)} \hat{x}$] and propagating along -$z$-direction, the electromagnetic wave equation in a plasma medium may be written as \cite{Panchenko},
%%%% Equation 3-2-3
\begin{equation} \label{eq3-2-3}
\partial_{tt} A - c^2 \left( \partial_{zz} A + \partial_{yy} A \right) + \omega_p^2 (z - v_p t) A = 0 .
\end{equation}
Here, $\omega_p (z - v_p t) = \sqrt{e^2 n ( z - v_p t ) / m \varepsilon_0 \gamma}$ is the plasma oscillation frequency. The electron density function of the mirror is expressed by $n (z - v_p t)$. For the relativistic flying mirror, the phase velocity, $v_p$ ($= \omega_p / k_p$), is the velocity of the mirror and it is almost the same as the speed of light, $c$. The $k_p$ is the plasma wavelength. For the flat mirror, we assume $\partial_{yy} A = 0$. From the Wentzel–Kramers–Brillouin (WKB) approximation, the electric field solution for Eq. \eqref{eq3-2-3} is given by, 
%%%% Equation 3-2-4
\begin{equation} \label{eq3-2-4}
E = E_s \left( \frac{\kappa}{\kappa_0} \right) \exp \left( -i \omega t - i \int_{-\infty}^{z} \kappa (z) d z \right) ,
\end{equation}
with the definition of $\kappa^2 (z) = k^2 + e^2 n (z - v_p t )/ m \varepsilon_0 c^2 \gamma$, where $k = \omega /c$. The $\kappa_0$ is the initial value of $\kappa$ at $z = -\infty$. The $E_s$ is the electric field strength of the counter-propagating source laser. The electron motion in the flying mirror can be described by
%%%% Equation 3-2-5
\begin{equation} \label{eq3-2-5}
\frac{d}{dt} \left( \gamma mc \vec{\beta} \right) = - e \left( \vec{E}_0 + \vec{E} + \vec{\beta} \times \vec{B} \right),
\end{equation}
where $\gamma$ is the Lorentz factor of the electron and $\vec{\beta}$ the electron velocity normalized to $c$. The $\vec{E}_0$ is given as $- \hat{z} E_0$ and responsible for the relativistic flying motion. The $\vec{E}$ (=$\hat{x} E$) is the electric field that results in the oscillatory motion of an electron. The $\vec{B}$ is then given by -$\hat{y} E /c$. The velocity, $\vec{\beta}$, of electron consists of two components as $\vec{\beta} = \hat{z} \beta_f + \hat{x} \beta_{os}$, where $\beta_f$ is $v_p/c \approx 1$ representing the relativistic flying motion and the $\beta_{os}$ is the oscillation velocity by the source pulse. By decomposing Eq. \eqref{eq3-2-5} into the flying and the oscillating motions, we may write
%%%% Equation 3-2-6
\begin{equation} \label{eq3-2-6}
\frac{d}{dt} \left( \gamma_M \beta_f \right) \approx \frac{e}{mc}E_0, \quad \textrm{and} \quad \frac{d}{dt} \left( \gamma_M \beta_{os} \right) \approx - \frac{e}{mc}E_s \left( \frac{\kappa}{\kappa_0} \right) \left( 1 - \beta_f\right) e^{- i \omega t - i \int_{-\infty}^z \kappa dz}.
\end{equation}
In Eq. \eqref{eq3-2-6}, it is assumed that $|\beta_f| \gg |\beta_{os}|$ and $\gamma \approx \gamma_M = 1/ \sqrt{1 - \beta_f^2}$. For the relativistic flying mirror with a constant $\gamma_M \gg 1$, the electron velocities can be re-written as,
%%%% Equation 3-2-7
\begin{equation} \label{eq3-2-7}
\beta_f \approx - \frac{e A_0}{\gamma_M mc}, \quad \textrm{and} \quad \beta_{os} \approx - i \frac{e E_s}{2 \gamma_M^3 mc \omega } e^{- i \omega t - i \int_{-\infty}^z \kappa dz},
\end{equation}
in the laboratory frame, and Eq. \eqref{eq3-2-7} becomes
%%%% Equation 3-2-8
\begin{equation} \label{eq3-2-8}
\beta'_f = 0, \quad \textrm{and} \quad \beta'_{os} \approx - i \frac{e E'_s}{ mc \omega' } e^{- i \omega' t' - i \int_{-\infty}^{z'} \kappa' dz'},
\end{equation}
in the boosted frame moving with a speed of $\beta_f$ along the +$z$-direction. Here, due to the Doppler effect, $E'_i \approx 2 \gamma_M E_i$ and $\omega' \approx 2 \gamma_M \omega$. 

%%%%%%%%%%% Figure 6
\begin{figure}%[b]
\centering
\includegraphics[width=1\columnwidth]{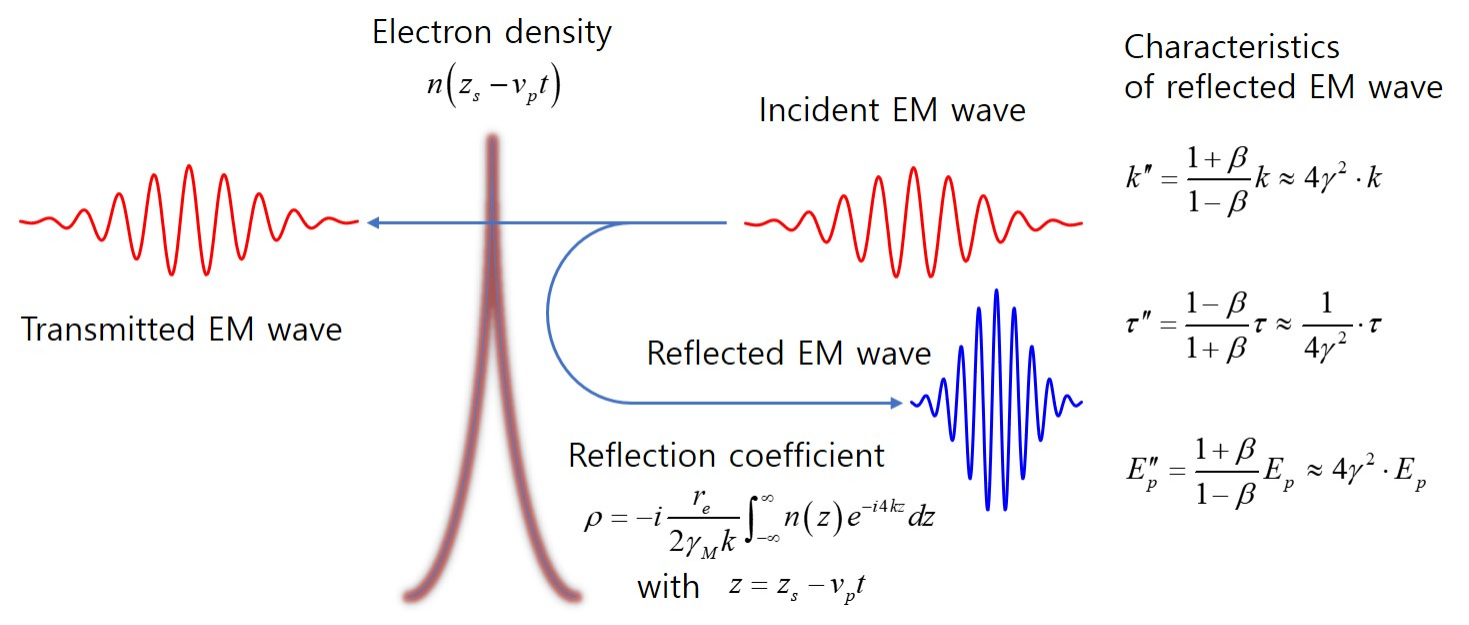}% Here is how to import EPS art \caption{\label{fig:epsart}
\caption{Reflection from the electron density layer.}
\end{figure}

The electric field radiated by an oscillating electron can be written by the Lienard-Wiechert retarded field as \cite{Jackson},
%%%% Equation 3-2-9
\begin{equation} \label{eq3-2-9}
\vec{E}' = - \frac{e}{4 \pi \varepsilon_0 c} \left[ \frac{\hat{n}'_s \times \hat{n}'_s \times \dot{\vec{\beta}}'_s }{ | \vec{r}'_o - \vec{r}'_s | } \right]_{t' = t'_r},
\end{equation}
where $\vec{r}'_o$ and $\vec{r}'_s$ stand for position vectors to the observation and source points from the origin, respectively. The distance vector, $\vec{r}'_{os}$, is defined by the polar ($\vartheta'$) and azimuthal ($\varphi'$) angles as,
%%%% Equation 3-2-11
\begin{equation} \label{eq3-2-11}
\vec{r}'_{os} \equiv \vec{r}'_o - \vec{r}'_s = r'_{os} \left( \hat{x}' \sin \vartheta' \cos \varphi' + \hat{y}' \sin \vartheta' \sin \varphi' + \hat{z}' \cos \vartheta' \right).
\end{equation}
The unit vector, $\hat{n}'_s = \left( \vec{r}'_o - \vec{r}'_s \right) / | \vec{r}'_o - \vec{r}'_s |$, is directed to the observation point from the source point. The retarded time, $t'_r$, is given by $t' - \left( 1 / \omega' \right) \int_{r'_s}^{r'_o} \kappa' d r'_{os}$. By inserting the retarded time into the time derivative of Eq. \eqref{eq3-2-8}, we obtain the expression for the acceleration as
%%%% Equation 3-2-10
\begin{equation} \label{eq3-2-10}
\dot{\beta}'_{os} |_{ret} \approx - \frac{e E'_s}{m c} \left( \frac{\kappa'}{\kappa'_0} \right) \exp \left( -i \omega' t' + i \int_{r'_s}^{r'_o} \kappa' d r'_{os} - i \int_{z'_i}^{z'_s} \kappa' dz' \right).
\end{equation}
After the straightforward calculation, the resultant field expression yields
%%%% Equation 3-2-12
\begin{eqnarray} \label{eq3-2-12}
\vec{E}' &\approx& r_e \frac{E'_s}{r'_{os}} \left( \frac{\kappa'}{\kappa'_0} \right) \exp \left( -i \omega' t' + \int_{r'_s}^{r'_o} \kappa' d r'_{os} - i \int_{z'_i}^{z'_s} \kappa' dz' \right) \nonumber \\
&&\times \left[ \hat{x}' \left( \sin^2 \vartheta' \sin^2 \varphi' + \cos^2 \vartheta' \right) + \hat{y}' \frac{\sin^2 \vartheta' \sin 2 \varphi'}{2} - \hat{z}' \frac{\sin 2 \vartheta' \sin \varphi'}{2} \right] ,
\end{eqnarray}
where $r_e$ is the classical electron radius defined as $e^2 / 4 \pi \varepsilon_0 m c^2$. 

We assume that electrons confined in a small region of the electron density are homogeneous in momentum. We further assume that the number of electrons should be large enough for the individual electron not to be considered an independent radiation emitter, but the Coulomb repulsion should be small enough during the interaction. Then, the total radiation electric field, $\vec{E}'_R$, is understood as the superposition of EM waves radiated from individual electrons within the relativistic flying mirror. The total radiation field can be calculated through the volume integral with the electron density function, $n' (z'_s)$, as, 
%%%% Equation 3-2-13
\begin{equation} \label{eq3-2-13}
\vec{E}'_R = \int n' (z'_s) \left[ \vec{E}' \right]_{t' = t'_r} dV',
\end{equation}
with $z'_s = \gamma_M (z - v_p t)$ and $d V' = {r'_{os}}^2 d r'_{os} d \cos \vartheta' d \varphi'$. In the weak reflection limit, we assume ${k'}^2 \gg 4 \pi e^2 n' (z') / m c^2$ \cite{Bulanov_Ref} so that $\kappa' \approx \kappa'_0 \approx k'$. The phase term in Eq. \eqref{eq3-2-12} can be approximated as
%%%% Equation 3-2-14
\begin{equation} \label{eq3-2-14}
- i \omega' t' + i \int_{r'_s}^{r'_o} \kappa' d r'_{os} - i \int_{z'_i}^{z'_s} \kappa' dz' \approx - i \omega' t' - i k' (z'_s - z'_i) + i k' \frac{z'_o - z'_s}{\cos \vartheta'}.
\end{equation}
By setting the initial phase shift as $k' z'_i = 0$ and considering an infinitely large relativistic flying mirror, the integration over $\varphi'$ yields 
%%%% Equation 3-2-15
\begin{equation} \label{eq3-2-15}
\vec{E}'_R \approx \hat{x}' \pi r_e E'_s \int r'_{os} dr'_s d \cos \vartheta' n' (z'_s) e^{- i \left( \omega' t' + k' z'_s - k' \frac{z'_o - z'_s}{\cos \vartheta'} \right)} \left( 1 + \cos^2 \vartheta' \right).
\end{equation}
Now, by using $z'_{os} = z'_o - z'_s$ and rewriting $r'_{os} d r'_s = z'_{os} d z'_s / \cos^2 \vartheta'$, the integration over $\vartheta'$, $\mathcal{I}_{\vartheta'}$, in Eq. \eqref{eq3-2-15} becomes
%%%% Equation 3-2-16
\begin{equation} \label{eq3-2-16}
\mathcal{I}_{\vartheta'} = \int_0^{\pi/2} d \cos \vartheta' \exp \left( i  \frac{k' z'_{os}}{\cos \vartheta'} \right) \frac{1 + \cos^2 \vartheta'}{\cos^2 \vartheta'} \approx \frac{1}{i k' z'_{os}} \exp (i k' z'_{os}).
\end{equation}
After inserting Eq. \eqref{eq3-2-16} into Eq. \eqref{eq3-2-15}, we obtain,
%%%% Equation 3-2-17
\begin{equation} \label{eq3-2-17}
\vec{E}'_R = - \hat{x}' i \pi  \frac{r_e E'_s}{k'} e^{-i \left( \omega' t' - k' z'_0 \right)} \int_{-\infty}^{\infty} dz'_s n' (z'_s) e^{- i 2 k' z'_s }.
\end{equation}
The integral on the right-hand side of Eq. \eqref{eq3-2-17} can be understood as the Fourier transform of the electron density function, $n'(z'_s)$. We obtain the reflection coefficient, $\rho'$, in the boosted frame by defining $\vec{E}'_R / \vec{E}' $ [=$\vec{E}'_R / \hat{x}E'_s e^{-i (\omega' t' - k' z'_o)}$] as, 
%%%% Equation 3-2-18
\begin{equation} \label{eq3-2-18}
\rho' =  -i \pi \frac{r_e}{k'} \mathcal{F} (2 k') \quad \textrm{with} \quad \mathcal{F} (2k') = \int_{-\infty}^{\infty} dz'_s n' (z'_s) e^{- i 2 k' z'_s },
\end{equation}
which is in agreement with the result obtained in \cite{Bulanov_Ref}. Since the $d z'_s n(z'_s)$ is Lorentz-invariant, the reflection coefficient in the laboratory frame can be written as,
%%%% Equation 3-2-19
\begin{equation} \label{eq3-2-19}
\rho =  -i \pi \frac{r_e}{ 2 \gamma_M k} \mathcal{F} (4k) \quad \textrm{with} \quad \mathcal{F} (4k) = \int_{-\infty}^{\infty} dz n (z) e^{- i 4 k z },
\end{equation}
where $z = z_s - v_p t$. In section 4, the reflection coefficient is calculated for a specific case, the cusp electron layer that appears in the laser-gas interaction, and its application to the SF-QED has been discussed.

\subsection{Electric and magnetic fields focused by the relativistic flying parabolic mirror}

Since the 4$\pi$-spherical focusing scheme can be applied in the boosted frame, it is convenient to calculate the field configuration in the boosted frame and then to Lorentz-transform the field to the laboratory frame. For the TM mode (radially-polarized laser), the focused field vectors in the boosted frame are expressed by \cite{Jeong_RFM},
%%%% Equation 3-3-1
\begin{equation} \label{eq3-3-1}
\left[ E'_{f,x}, E'_{f,y}, E'_{f,z} \right] = i E'_p (\omega') a (\rho', \theta' ; \omega') e^{i \omega' t'} 
\left[ \cos \theta' \cos \phi', \cos \theta' \sin \phi', - \sin \theta' \right]
\end{equation}
and
%%%% Equation 3-3-2
\begin{equation} \label{eq3-3-2}
\left[ B'_{f,x} , B'_{f,y} , B'_{f,z} \right] = - B'_p (\omega') b (\rho', \theta' ; \omega') e^{i \omega' t'} 
\left[ \sin \phi' , - \cos \phi' , 0 \right] ,
\end{equation}
with expressions of $a(\rho', \theta'; \omega')$ and $b(\rho', \theta'; \omega')$ redefined in the boosted frame as,
%%%% Equation 3-3-3
\begin{equation} \label{eq3-3-3}
a(\rho', \theta'; \omega') \approx j_0 \left( \frac{\omega'}{c} \rho'\right) + \frac{5}{2^3} j_2 \left( \frac{\omega'}{c} \rho'\right) P_2 (\cos \theta'),
\end{equation}
and
%%%% Equation 3-3-4
\begin{equation} \label{eq3-3-4}
b(\rho', \theta'; \omega') \approx \frac{4}{\pi} j_1 \left( \frac{\omega'}{c} \rho'\right) P_1^1 (\cos \theta').
\end{equation}

For the laser pulse reflected and focused by the RFPM, the general expressions for the spatiotemporal field distribution in the laboratory frame are given by,
%%%% Equation 3-3-5
\begin{equation} \label{eq3-3-5}
\vec{E}''_f = \gamma_M \frac{1+ \beta_f}{ 1 - \beta_f} \sqrt{\frac{3 \pi}{c \epsilon_0}} \frac{\pi \omega_0 r_0 \sqrt{I_s}}{4c} 
\left[ f_x , f_y , f_z \right],
\end{equation}
and
%%%% Equation 3-3-6
\begin{equation} \label{eq3-3-6}
\vec{B}''_f = \frac{\gamma_M}{c} \frac{1+ \beta_f}{ 1 - \beta_f} \sqrt{\frac{3 \pi}{c \epsilon_0}} \frac{\pi \omega_0 r_0 \sqrt{I_s}}{4c} 
\left[ h_x , h_y , 0 \right],
\end{equation}
where $I_s$ is the intensity of the source laser pulse. The vector components in Eqs. \eqref{eq3-3-5} and \eqref{eq3-3-6} are given as,
%%%% Equation 3-3-7
\begin{eqnarray} \label{eq3-3-7}
f_x &=& \left[ -j_0 \left( \frac{\omega'_0 R}{c} \right) \sin (\omega'_0 T) \Upsilon_1  + \beta_f j_1 \left( \frac{\omega'_0 R}{c} \right) \cos (\omega'_0 T) \Upsilon_2 \right] \cos \phi , \\
f_y &=& \left[ -j_0 \left( \frac{\omega'_0 R}{c} \right) \sin (\omega'_0 T) \Upsilon_1  + \beta_f j_1 \left( \frac{\omega'_0 R}{c} \right) \cos (\omega'_0 T) \Upsilon_2 \right] \sin \phi , \\
f_z &=& \frac{1}{\gamma_M} j_0 \left( \frac{\omega'_0 R}{c} \right) \sin (\omega'_0 T) \Upsilon_2  , \\
h_x &=& - \left[ j_1 \left( \frac{\omega'_0 R}{c} \right) \cos (\omega'_0 T) \Upsilon_2 - \beta_f j_0 \left( \frac{\omega'_0 R}{c} \right) \sin (\omega'_0 T) \Upsilon_1 \right] \sin \phi, \\
\textrm{and} \nonumber \\
h_y &=& \left[ j_1 \left( \frac{\omega'_0 R}{c} \right) \cos (\omega'_0 T) \Upsilon_2 - \beta_f j_0 \left( \frac{\omega'_0 R}{c} \right) \sin (\omega'_0 T) \Upsilon_1 \right] \cos \phi .
\end{eqnarray}
Here, $\omega'_0$ is the center frequency of the source laser pulse in the boosted frame. The time-like, $T$, and space-like, $R$, functions are expressed as,
%%%% Equation 3-3-10
\begin{equation} \label{eq3-3-10}
T = T(t,\rho) = \frac{t - (\rho/c) \beta_f \cos \theta}{\gamma_M \left( 1 - \beta_f^2 \cos^2 \theta\right)},
\end{equation}
and
%%%% Equation 3-3-11
\begin{equation} \label{eq3-3-11}
R = R(t,\rho) = \frac{\rho - ct \beta_f \cos \theta}{\gamma_M \left( 1 - \beta_f^2 \cos^2 \theta\right)}.
\end{equation}
Here, $t$ is the time, and $\rho$ the radial position in the laboratory frame. The envelop functions, $\Upsilon_1$ and $\Upsilon_2$, are defined as,
%%%% Equation 3-3-8
\begin{equation} \label{eq3-3-8}
\Upsilon_1 = \frac{1}{2} \left\{ \frac{\cos \theta + \beta_f}{1+ \beta_f \cos \theta} \exp \left[ - \frac{\Delta {\omega'}^2}{4} \left( T + \frac{R}{c} \right)^2 \right] + \frac{\cos \theta - \beta_f}{1- \beta_f \cos \theta} \exp \left[ - \frac{\Delta {\omega'}^2}{4} \left( T - \frac{R}{c} \right)^2 \right] \right\},
\end{equation}
and
%%%% Equation 3-3-9
\begin{equation} \label{eq3-3-9}
\Upsilon_2 = \frac{1}{2} \left\{ \frac{\sin \theta / \gamma_M}{1+ \beta_f \cos \theta} \exp \left[ - \frac{\Delta {\omega'}^2}{4} \left( T + \frac{R}{c} \right)^2 \right] + \frac{\sin \theta / \gamma_M}{1- \beta_f \cos \theta} \exp \left[ - \frac{\Delta {\omega'}^2}{4} \left( T - \frac{R}{c} \right)^2 \right] \right\},
\end{equation}
with the spectral bandwidth, $\Delta \omega'$, of the source laser pulse. Figure 7 shows the evolution of the spatio-temporal distribution of the electric field reflected and focused by the RFPM. The general (backward curved) feature shows a good agreement between the numerical \cite{SVBulanov_RFM} and the analytical calculations of the focused field by the relativistic flying mirror. 

%%%%%%%%%%% Figure 7
\begin{figure}[b]
\centering
\includegraphics[width=1\columnwidth]{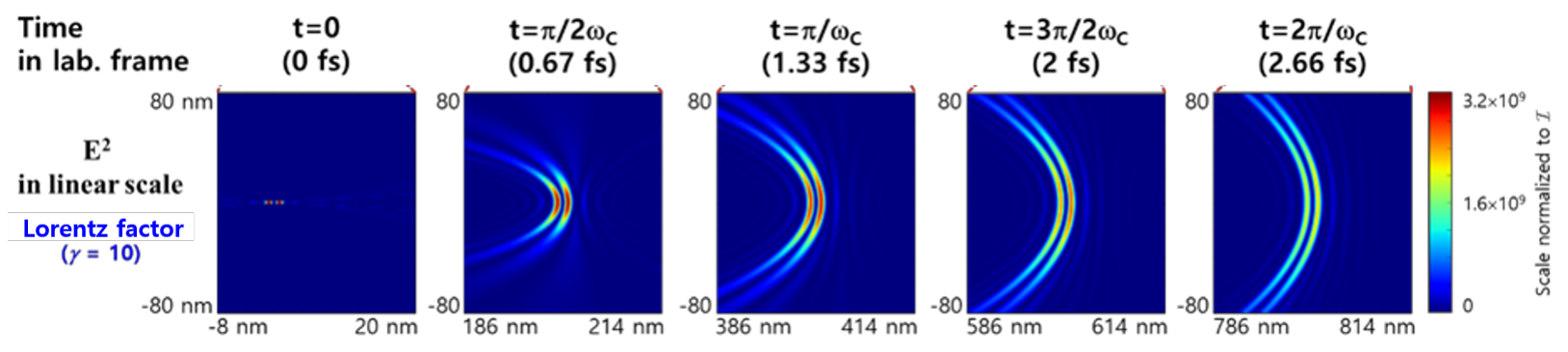}% Here is how to import EPS art \caption{\label{fig:epsart}
\caption{The evolution of the spatiotemporal distribution of the electric field reflected and focused by the relativistic flying parabolic mirror. Modified from Fig. 3 in \cite{Jeong_RFM}}
\end{figure}

In the relativistic limit of $\gamma_M \gg 1$, $\Upsilon_1 \sim 1 \gg \Upsilon_2$. The peak value for $\Upsilon_1$ depends on the $\Delta \omega' / \omega'_0$, and the calculation shows that the peak value of $\Upsilon_1$ is about 0.94 for $\Delta \omega' / \omega'_0 = 0.075$ at $\theta = 0$. From Eq. \eqref{eq3-3-5}, it is obvious that the field enhancement factor is approximated as 
%%%% Equation 3-3-12
\begin{equation} \label{eq3-3-12}
\gamma_M \frac{1+ \beta}{ 1 - \beta} \sqrt{\frac{3 \pi}{c \epsilon_0}} \frac{\pi \omega_0 r_0 \sqrt{I_s}}{4c} \approx \pi^2 \sqrt{\frac{3 \pi}{2}} \gamma_M^3 \left( \frac{2r_0}{\lambda_0} \right) E_s. 
\end{equation}
For the case of  $\Delta \omega' / \omega'_0 = 0.075$, by multiplying 0.94 to Eq. \eqref{eq3-3-12}, the peak field strength can be approximated as
%%%% Equation 3-3-13
\begin{equation} \label{eq3-3-13}
E''_f \approx 20.1 \gamma_M^3 \left( \frac{2r_0}{\lambda_0} \right) E_s. 
\end{equation}
The final field strength should be calculated by multiplying the reflection coefficient expressed in Eq. \eqref{eq3-2-17} to Eq. \eqref{eq3-3-13}. In a realistic case, the shape of the relativistic flying mirror may deviate from the ideal paraboloidal one. One of the main reasons for deviation might be the wavefront aberration of the driver laser pulse which is not considered in this study. The deviation can be imprinted as the wavefront aberration to the source laser pulse reflected from the relativistic flying mirror, and it reduces the focused intensity of a source laser pulse. For the normal focusing case with a small amount of wavefront aberration, the intensity reduction can be estimated by the Strehl ratio given by 
%%%% Equation 3-3-14
\begin{equation} \label{eq3-3-14}
\exp \left(-\sigma_{\Phi}^2 \right), 
\end{equation}
where $\sigma_{\Phi}$ is the variance of the wavefront aberration in \cite{Mahajan}. Considering a stereographic projection from the 3D spherical surface to the 2D plane \cite{Pedoe}, the approach using Eq. \eqref{eq3-3-14} might be still applicable to assess the intensity reduction with a projected wavefront aberration.

\section{Application to the strong field quantum electrodynamics: collision with \texorpdfstring{$\gamma$-}{gamma}photons}
Since the source laser intensity reflected and focused by the RFPM can reach a super-strong field, many interesting ideas using the RFPM have been proposed in \cite{Koga,Chen,Kando_2022}. And it is also interesting to employ such a super-strong field strength for the SF-QED study by colliding it with $\gamma$-photons. The RFPM can be formed via the laser-gas interaction as described in \cite{Kando}, and the high-energy $\gamma$-photons are produced via the laser-solid interactions \cite{Nakamura,Ji,Wang,Hadjisolomou1,Hadjisolomou2,Hadjisolomou3}. First, let us estimate how high intensity can be obtained by the RFPM. Since the electron distribution in the vicinity of the wave-breaking threshold in a cold plasma is known as the cusp electron layer, we consider a cusp electron density for the relativistic flying mirror. The temperature of the electron layer broadens the electron density profile and reduces the reflectance of the EM radiation from the layer. Such effect of the temperature on the electron density layer and its relevant reflectance have been studied for the cusp electron density in \cite{Bulanov_Thermal,Bulanov_Ref}. Note that some proposals for the relativistic flying mirror use the up-ramp density profile to generate an ultrathin and strongly over-dense relativistic electron sheet with a higher $\gamma$-factor \cite{Li_PRL, Li_APL}. 

In the laboratory frame, the electron density for the cusp layer is approximated as \cite{Panchenko},
%%%% Equation 4-1-1
\begin{equation} \label{eq4-1-1}
n (z_s, t) \approx \left( \frac{2}{9} \right)^{1/3} \frac{ n_0 \left( 1 + \eta_d^2 \right)^{1/6} \gamma_M }{ \left( k_{p0} l \right)^{2/3} \left[ 1 + (z_s - \beta_f c t )^2 /l^2 \right]^{1/3}} ,
\end{equation}
where $k_{p0}$ is $\omega_{p0} / v_p$ with the plasma frequency, $\omega_{p0} = \sqrt{n_0 e / m \varepsilon_0 \gamma_M}$. The $\eta_d$ is the field strength of the driver laser pulse. The  $l$ is the characteristic width of the cusp so that the $k_{p0} l$ defines a sharpness parameter on how close the plasma wave is to the wave-breaking limit \cite{Bulanov_AM}. The electron density in the boosted frame may be written as,
%%%% Equation 4-1-2
\begin{equation} \label{eq4-1-2}
n' (z'_s) \approx \left( \frac{2}{9} \right)^{1/3} \frac{ n'_0 \left( 1 + \eta_d^2 \right)^{1/6} \gamma_M^{4/3} }{ \left( k'_{p0} l' \right)^{2/3} \left[ 1 + (z'_s /l' )^2 \right]^{1/3}} .
\end{equation}
By plugging the density function into Eq. \eqref{eq3-2-18}, we obtain the reflection coefficient for the cusp electron layer as, 
%%%% Equation 4-1-3
\begin{equation} \label{eq4-1-3}
\rho' = -i \sqrt{\pi} \left( \frac{2}{9} \right)^{1/3} \Gamma \left( \frac{2}{3} \right)  r_e \frac{ n'_0 {l'}^{1/6} }{{k'}^{7/6} {k'_{p0}}^{2/3}} \left( 1 + \eta_d^2 \right)^{1/6} \gamma_M^{4/3} K_{1/6} (2 k' l' ) ,
\end{equation}
with the help of the integral identity of
%%%% Equation 4-1-4
\begin{equation} \label{eq4-1-4}
\int_0^{\infty} \left( x^2 + a^2 \right)^{\nu - 1/2} \cos (bx) dx = \frac{1}{\sqrt{\pi}} \left( \frac{2a}{b} \right)^{\nu} \cos (\nu \pi) \Gamma (\nu + 1/2) K_{-\nu} (ab) .
\end{equation}
In Eq. \eqref{eq4-1-3}, the electron density is eliminated by using the relationship of $n'_0 = {k'_{p0}}^2 / 4 \pi r_e$, and the reflection coefficient may be rewritten as
%%%% Equation 4-1-5
\begin{equation} \label{eq4-1-5}
\rho' = -i \frac{(2/9)^{1/3}}{4\sqrt{\pi}} \Gamma \left( \frac{2}{3} \right)  \frac{ {k'_{p0}}^{4/3} l'^{1/6} }{ k'^{7/6} } \gamma_M^{4/3} \left( 1 + \eta_d^2 \right)^{1/6} K_{1/6} (2 k' l' )  .
\end{equation}
Finally, in the laboratory frame, we have
%%%% Equation 4-1-6
\begin{subequations} \label{eq4-1-6}
\begin{align}
\rho &= - i \frac{(2/9)^{1/3}}{8 \cdot 2^{1/6} \sqrt{\pi}} \Gamma \left( \frac{2}{3} \right) \frac{1}{\gamma_M^{1/3}} \frac{ k_{p0}^{4/3} l^{1/6} }{k^{7/6}} \left( 1 + \eta_d^2 \right)^{1/6} K_{1/6} (4 k l )  \\
&\approx -i 0.051 \frac{1}{\gamma_M^{1/3}} \frac{ k_{p0}^{4/3} l^{1/6} }{k^{7/6}} \left( 1 + \eta_d^2 \right)^{1/6} K_{1/6} (4
k l ).
\end{align}
\end{subequations}

The modified Bessel function of the second kind, $K_{1/6} (\cdot)$, is an exponential decay function, but in the wave-braking limit (where $0 < 4 k l \ll \sqrt{\nu}$) the modified Bessel function of the second kind has an asymptotic form of $K_{\nu} (x) \sim \left( \Gamma (\nu) /2 \right) \left( 2/x \right)^{\nu}$ \cite{AbramowiczAndStegun}. Thus, in the wave-breaking limit, the reflection coefficient is given as,
%%%% Equation 4-1-7
\begin{subequations} \label{eq4-1-7}
\begin{align}
\rho_{\textrm{wb}} &\approx -i \frac{(2/9)^{1/3}}{16 \cdot 2^{1/3} \sqrt{\pi}} \Gamma \left( \frac{2}{3} \right) \Gamma \left( \frac{1}{6} \right) \gamma_M^{-1/3} \left( \frac{ k_{p0} }{k} \right)^{4/3} \left( 1 + \eta_d^2 \right)^{1/6} \\
&\approx -i 0.128 \gamma_M^{-1/3} \left( \frac{ k_{p0} }{k} \right)^{4/3} \left( 1 + \eta_d^2 \right)^{1/6} .
\end{align}
\end{subequations}
This formula provides the maximum reflection coefficient obtained by the relativistic flying mirror. Figure 8(a) and 8(b) show the characteristic width of the cusp electron layer and the reflection coefficient, $|\rho |$, with respect to the sharpness parameter, $k_{p0} l$. The wavelength of 0.8 $\mu$m is selected assuming the use of Ti:sapphire laser and the electron density, $n_0$, varies from 1$\times$10$^{18}$ cm$^{-3}$ to 5$\times$10$^{18}$ cm$^{-3}$. The field strength for the driver laser pulse, $\eta_d$, is 2. The Lorentz $\gamma$-factor of the mirror, $\gamma_M$, is a function of electron density given as $\sqrt{n_{cr} / n_0}$ \cite{Yoffe,Gupta}, yielding the mirror speed of >0.99$c$ in the density range. For the use of 4 PW laser pulse as a driver, the radius of the focal spot for $\eta_d = 2$ is about 120 $\mu$m. So the corresponding radius for the RFPM is also about 120 $\mu$m. As the sharpness parameter, $k_{p0} l$, decreases from 0.05 to the wave-breaking limit ($k_p l \to 0$),
the reflection coefficients, $|\rho|$, increase from 2.8$\times$10$^{-12}$ to 3.2$\times$10$^{-4}$ for $n_0 = 1 \times 10^{18}$ cm$^{-3}$, from 1.0$\times$10$^{-8}$ to 8.3$\times$10$^{-4}$ for $n_0 = 3 \times 10^{18}$ cm$^{-3}$, from 1.5$\times$10$^{-7}$ to 1.3$\times$10$^{-3}$ for $n_0 = 5 \times 10^{18}$ cm$^{-3}$, respectively.

%%%%%%%%%%% Figure 8
\begin{figure}%[b]
\centering
\includegraphics[width=1\columnwidth]{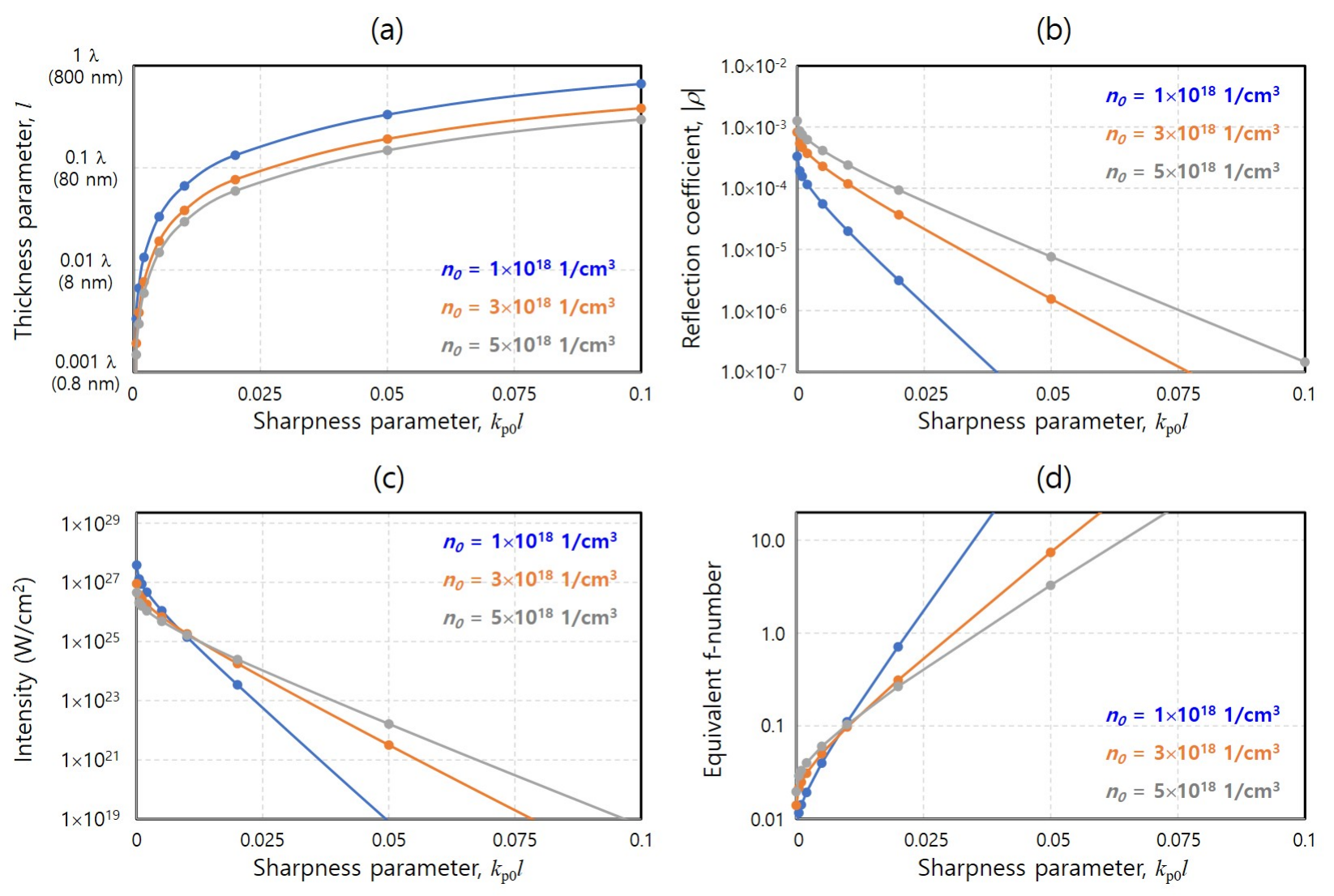}% Here is how to import EPS art \caption{\label{fig:epsart}
\caption{(a) Characteristic width of the cusp electron layer with respect to the sharpness parameter. (b) Reflection coefficient for the cusp electron layer with respect to the sharpness parameter. (c) Reflected and focused intensity of the source laser pulse with respect to the sharpness parameter. (d) Equivalent geometric f-number for reaching the reflected and focused source laser intensity.}
\end{figure}

The intensity enhancement for the source laser pulse is shown in Fig. 8(c). A source laser pulse with an intensity of 1.8$\times 10^{17}$ W/cm$^2$ ($\eta \approx 0.29$) is considered for this calculation. An ideal reflectance condition (i.e., $\mathcal{R}$ is independent of the incident angle) is assumed for this calculation. Depending on the gas density, $n_0$, the highest peak intensities reach at 3.9$\times$10$^{27}$ W/cm$^2$ for $n_0 = 1 \times 10^{18}$ cm$^{-3}$, 9.1$\times$10$^{26}$ W/cm$^2$ for $n_0 = 3 \times 10^{18}$ cm$^{-3}$, and 4.6$\times$10$^{26}$ W/cm$^2$ for $n_0 = 5 \times 10^{18}$ cm$^{-3}$, respectively. It is obvious that a source laser pulse reflected by a smaller sharpness parameter surpasses the intensity obtained by the optically-focusing scheme \cite{Jeong_4pi}. The equivalent f-numbers, $f/\#$, as the geometrical values ($D/f$) that provide the same laser intensities with the 4 PW laser pulse are calculated and shown in Fig. 8(d).     

Due to the super-strong field generated by the RFPM, it is interesting to estimate the quantum nonlinearity parameter obtained with high-energy $\gamma$-photons. Figure 9(a) shows the schematic diagram for colliding high-energy $\gamma$-photon with the intensified laser field by the RFPM. The high-energy $\gamma$-photon is generated from a low-Z solid target irradiated by an intense ($\sim$10$^{24}$ W/cm$^2$) laser pulse. 

%%%%%%%%%%% Figure 9
\begin{figure}[b]
\centering
\includegraphics[width=1\columnwidth]{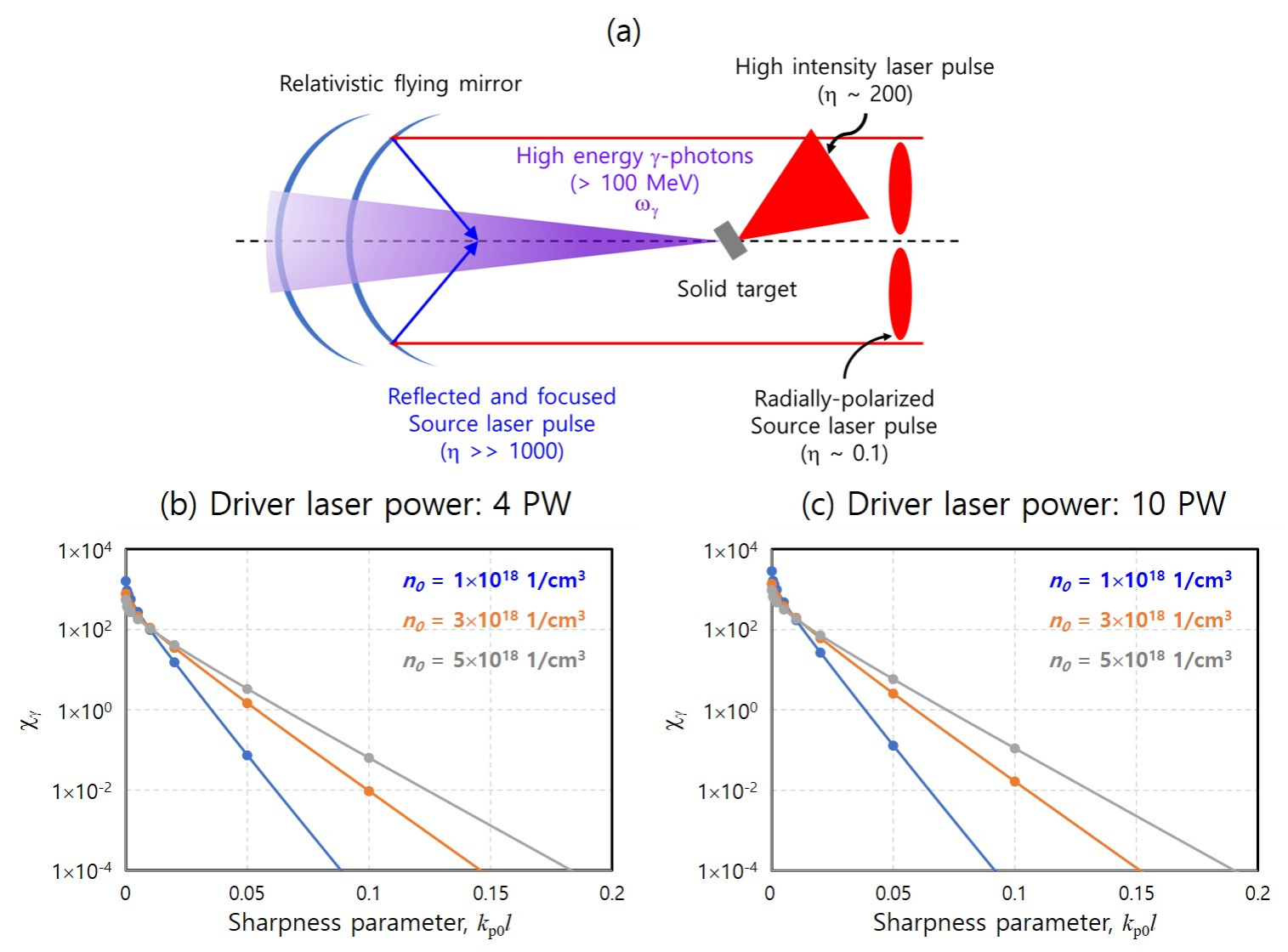}% Here is how to import EPS art \caption{\label{fig:epsart}
\caption{(a) Collision scheme between the high-energy $\gamma$-photons and the ultra-strong source laser pulse reflected and focused by the RFPM. A low-Z metallic target is irradiated by another fs high-power (10 PW) laser to generate $\gamma$-photons. Quantum nonlinearity parameters (b) when colliding the 900 MeV $\gamma$-photon and (c) when colliding the 1 GeV $\gamma$-photon with the source laser pulse ($\eta$ = 0.29) reflected and focused by a RFPM with a $\gamma$-factor of $\sim$22. The RFPMs are driven by 4 PW Ti:sapphire laser pulse with a spot size of 240 $\mu$m in (b) and by 10 PW Ti:sapphire laser pulse with a spot size of 380 $\mu$m in (c).}
\end{figure}

According to \cite{Hadjisolomou2}, the maximum energy for the high-energy $\gamma$-photon extends to 900 MeV. The high-energy $\gamma$-photons have a two-lobes angular distribution structure. So, the target is rotated by $\sim$35$^{\circ}$ to direct the maximum $\gamma$-photon fluence to the high field strength. The emitted $\gamma$-photons propagate to the relativistic flying focus for the head-on collision. From Eqs. \eqref{eq1-3}, \eqref{eq3-3-13}, and \eqref{eq4-1-6}, the quantum nonlinearity parameter, $\chi_{\gamma}$, for the counter-propagating case may be written as,
%%%% Equation 4-1-8
\begin{equation} \label{eq4-1-8}
\chi_{\gamma} \approx 2  \frac{\omega_{\gamma}}{m} \gamma_M^{8/3} \frac{ k_p^{4/3} l^{1/6} }{k^{7/6}}  \left( 1 + \eta_d^2 \right)^{1/6} K_{1/6} \left( 4 k l \right) \frac{2 w_0}{\lambda} \frac{ E_s}{E_c},
\end{equation}
or in the wave-breaking limit
%%%% Equation 4-1-9
\begin{equation} \label{eq4-1-9}
\chi_{\gamma,\textrm{wb}} \approx 5.1 \frac{\omega_{\gamma}}{m} \gamma_M^{8/3} \left( \frac{ k_p }{k} \right)^{4/3} \left( 1 + \eta_0^2 \right)^{1/6} \frac{2 w_0}{\lambda} \frac{ E_s}{E_c} .
\end{equation}

Figures 9(b) and 9(c) show the quantum nonlinearity parameter, $\chi_{\gamma}$, when high-energy $\gamma$-photons collide with the reflected and focused source laser field at two different driver laser power/$\gamma$ photon energy combinations. For the 10 PW driver laser power/1 GeV $\gamma$-photon case, the highest $\chi_{\gamma}$ increases up to $\sim$2850 at $n_0 = 1 \times 10^{18}$ cm$^{-3}$. This value is well above $\alpha^{-3/2}$, which is the condition for the non-perturbative SF-QED study \cite{Narozhny}.

\section{Conclusion}
The generation of the super-strong field by the relativistic flying mirror has been reviewed. The peak field strength has been calculated for the cusp electron layer model which is the electron density appearing when an intense laser pulse propagates in the plasma medium. The calculation shows that a weakly relativistic source laser pulse ($\eta = 0.29$) reflected and focused by an ideal cusp electron layer with a constant reflectance can be intensified to a field strength of $\eta > 2000$ in the wave-breaking limit. This intensified source laser field exceeds the peak laser intensity that can be obtained through the optical-focusing scheme. Such a super-strong field can be applied to the strong-field quantum electrodynamics by colliding it with high-energy $\gamma$-photons. The peak value of quantum nonlinearity parameter easily exceeds the unity, and it can reach well above $\sim$1600 ($\alpha^{-3/2}$) when the 1-GeV $\gamma$-photon collides with a super-strong source laser field reflected and focused by the relativistic flying parabolic mirror driven by a 10 PW laser pulse. The 10 PW lasers commissioned or being commissioned in ELI facilities can be combined with a synergic approach that can enhance the effective field strength and might be good tools for testing strong-field quantum electrodynamics in perturbative and non-perturbative regimes. 

This review provides the theoretical framework for implementing the relativistic flying parabolic mirror for the test of strong-field quantum electrodynamics via the synergic approach. The approach is based on an ideal paraboloidal mirror profile. Thus, it delivers the theoretical limit for the field strength and the quantum nonlinearity parameter. In a realistic case, many factors, including the realistic shape of the relativistic flying mirror, incident-angle dependent reflectance, and so on, will affect and reduce the parameter from the theoretical limit. However, the synergic approach using the relativistic flying mirror can still provide unprecedented physical parameters in the laboratory for interesting fundamental science.

\backmatter

%\bmhead{Supplementary information}

%If your article has accompanying supplementary file/s please state so here. 

%Authors reporting data from electrophoretic gels and blots should supply the full unprocessed scans for key as part of their Supplementary information. This may be requested by the editorial team/s if it is missing.

%Please refer to Journal-level guidance for any specific requirements.

\bmhead{Acknowledgments}
This work was supported by the project
“Advanced research using high intensity laser
produced photons and particles” (ADONIS)
(CZ.02.1.01/0.0/0.0/16\_019/0000789) from the European
Regional Development Fund.
%Acknowledgments are not compulsory. Where included they should be brief. Grant or contribution numbers may be acknowledged.

%Please refer to Journal-level guidance for any specific requirements.

\section*{Declarations}
%Some journals require declarations to be submitted in a standardised format. Please check the Instructions for Authors of the journal to which you are submitting to see if you need to complete this section. If yes, your manuscript must contain the following sections under the heading `Declarations':

\begin{itemize}
\item Conflict of interest: There is no conflict of interest.
%\item Ethics approval 
%\item Consent for publication
%\item Availability of data and materials
%\item Authors' contributions
\end{itemize}

\noindent

\bibliography{sn-bibliography}% common bib file

\end{document}